\documentclass[aps, pra, twocolumn, amsfonts, amsmath, amssymb, superscriptaddress, floatfix]{revtex4-2}

\usepackage[utf8x]{inputenc}  
\usepackage[T1]{fontenc} 
\usepackage[english]{babel}
\usepackage{graphicx}  
\usepackage{mathbbol} 
\usepackage{kantlipsum}
\usepackage{float}
\usepackage{tabularx}
\usepackage{booktabs}
\usepackage{siunitx} 
\usepackage[colorlinks,hyperindex]{hyperref}
\hypersetup{colorlinks, citecolor=blue, linkcolor=blue, urlcolor=blue}
\usepackage[caption=false]{subfig} 
\usepackage{mathtools}  
\usepackage{amssymb}
\usepackage[noformat]{mattens} 
\usepackage[dvipsnames,svgnames]{xcolor}
\usepackage{mathrsfs} 

\renewcommand{\a}{\hat{a}}
\renewcommand{\b}{\hat{b}}
\renewcommand{\c}{\hat{c}}
\newcommand{\w}{\hat{w}}
\newcommand{\x}{\hat{x}}
\renewcommand{\H}{\hat{H}}
\newcommand{\C}{\mathcal{C}}
\newcommand{\D}{\mathcal{D}}
\newcommand{\Ds}{\mathcal{D}_\mathrm{s}}
\newcommand{\G}{\mathcal{G}}
\renewcommand{\L}{\mathcal{L}}
\newcommand{\Lk}{\hat{L}}
\renewcommand{\O}{\hat{ \scalebox{0.85}{$\mathcal{O}$}}}
\newcommand{\R}{{\scriptscriptstyle \mathrm{R}}}
\renewcommand{\S}{{\scriptscriptstyle \mathrm{S}}}
\newcommand{\tot}{{\scriptscriptstyle \mathrm{T}}}

\newcommand{\Tr}{\mathrm{Tr}}

\newcommand{\dd}{\mathrm{d}}
\renewcommand{\Re}{\mathrm{Re}}
\renewcommand{\Im}{\mathrm{Im}}
\newcommand{\dv}[2]{\frac{\dd #1}{\dd #2}}

\makeatletter
\def\NAT@def@citea{\def\@citea{\NAT@separator}}
\makeatother

\usepackage[color=green!60,textwidth=15mm]{todonotes}

\usepackage[normalem]{ulem} 

\usepackage{changes}
\setaddedmarkup{\textcolor{blue!80}{#1}}
\setdeletedmarkup{\textcolor{red!80}{\sout{#1}}}

\begin{document}

\title[Non-diagonal Lindblad master equations in quantum reservoir engineering]{Non-diagonal Lindblad master equations in quantum reservoir engineering}
\author{D. N. Bernal-Garc{\'\i}a}
\email{d.bernalgarcia@griffith.edu.au}
\affiliation{School of Engineering and Information Technology, UNSW Canberra, ACT 2600, Australia}
\affiliation{Grupo de Superconductividad y Nanotecnolog{\'\i}a, Universidad Nacional de Colombia, Ciudad Universitaria, K. 45 No. 26-85, Bogot{\'a} D.C., Colombia}
\affiliation{Centre for Quantum Dynamics, Griffith University, Yuggera Country, Brisbane, Queensland 4111, Australia}

\author{L. Huang}
\author{A. E. Miroshnichenko}
\author{M. J. Woolley}
\email{m.woolley@unsw.edu.au}
\affiliation{School of Engineering and Information Technology, UNSW Canberra, ACT 2600, Australia}

\date{\today}


\begin{abstract}
Reservoir engineering has proven to be a practical approach to control open quantum systems, preserving quantum coherence by appropriately manipulating the reservoir and system-reservoir interactions.
In this context, for systems comprised of different parts, it is common to describe the dynamics of a subsystem of interest by performing an adiabatic elimination of the remaining components of the system.
This procedure often leads to an effective master equation for the subsystem that is not in the diagonal form of the Gorini-Kossakowski-Lindblad-Sudarshan master equation (here called diagonal Lindblad form).
Instead, it has a more general structure (here called non-diagonal Lindblad form), which explicitly reveals the dissipative coupling between the various components of the subsystem.
In this work, we present a set of dynamical equations for the first and second moments of the canonical variables for linear Gaussian systems, bosonic and fermionic, described by non-diagonal Lindblad master equations.
Our method is efficient and allows one to obtain analytical solutions for the steady state.
We supplement our findings with a review of covariance matrix methods, focusing on those related to the measurement of entanglement. Notably, our exploration yields a surprising byproduct: the Duan criterion, commonly applied to bosonic systems for verification of entanglement, is found to be equally valid for fermionic systems.
We conclude with a practical example, where we revisit two-mode mechanical entanglement in an optomechanical setup. Our approach, which employs adiabatic elimination for systems governed by time-dependent Hamiltonians, opens the door to examine physical regimes that have not been explored before.
\end{abstract}

\maketitle 
%

%
%
%
%

\section{Introduction}
\label{sec:introduction}

In the quantum mechanics of open systems, one often comes across the question: Is the interaction with the environment a friend or a foe? And although the naïve answer is that noise and dissipation are enemies of the pure quantum behavior of nature, paradoxically, it is now well known that it is possible to engineer reservoirs and system-reservoir interactions in such a way that quantum coherence is preserved through dissipative dynamics.
This quantum reservoir engineering has been the subject of extensive research, especially in the context of optical and atomic systems~\cite{Cirac1993, Poyatos1996, Lutkenhaus1998, Plenio2002, Muschik2011,Krauter2011,Lin2013,Kienzler2015}.
Further, the importance of reservoir engineering has also been demonstrated in systems of superconducting qubits~\cite{Murch2012, Shankar2013}.
In recent years, the work being done in quantum reservoir engineering of optomechanical systems has gained considerable interest due to its potential application both in the investigation of the fundamental limits of quantum mechanics and in the development of quantum technologies.
It is worth highlighting the preparation of steady-state mechanical squeezed states~\cite{Kronwald2013,Wollman2015,Lecocq2015,Pirkkalainen2015,Lei2016}, and the realization of stabilized entanglement between macroscopic mechanical resonators~\cite{Wang2013,Woolley2014,Ockeloen-Korppi2018,MercierdeLepinay2021}.
\begin{figure}[b]
    \centering
    \includegraphics[width=\linewidth]{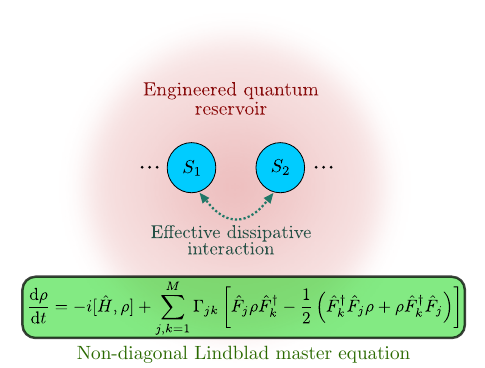}
    \caption{Master equations in non-diagonal Lindblad form \eqref{eq:non-diagonal_lindblad_form} appear naturally in the description of systems weakly interacting with an engineered quantum reservoir, where they often explicitly reveal the effective dissipative coupling between non-interacting subsystems.
    \label{fig:graphic_abstract}}
\end{figure}
In the theory of open quantum systems, the non-unitary dynamics of the density matrix $\rho(t)$ representing the physical state of a system interacting with its environment is, within the Markov approximation, generally described by the celebrated Gorini-Kossakowski-Lindblad-Sudarshan~(GKLS) master equation~\cite{Gorini1976,Lindblad1976} (here we refer to as diagonal Lindblad form):
\begin{align}
    &\dv{\rho}{t} = -i [\H, \rho] + \sum\limits_{j=1}^M \gamma_j \left[ \Lk_j \rho \Lk_j^\dagger - \frac{1}{2} \left( \Lk_j^{\dagger} \Lk_j \rho + \rho \Lk_j^{\dagger} \Lk_j \right) \right], \label{eq:diagonal_lindblad_form} 
\end{align}
where $\H$ is the system Hamiltonian, and the $\gamma_j \in \mathbb{R}$ correspond to the rates associated with each decoherence operator $\Lk_j$  ($j=1, \dots, M$).
Nonetheless, in the context of reservoir engineering, one often arrives at master equations with a more general structure, which enables coupling between different decoherence channels (here we call this non-diagonal Lindblad form):
\begin{align}
    \dv{\rho}{t} =&  -i [\H, \rho] \nonumber \\
    & + \sum\limits_{j,k=1}^M \Gamma_{j k} \left[ \hat{F}_j \rho \hat{F}_k^\dagger - \frac{1}{2} \left( \hat{F}_k^\dagger \hat{F}_j \rho + \rho \hat{F}_k^\dagger \hat{F}_j \right) \right], \label{eq:non-diagonal_lindblad_form} 
\end{align}
where the $\Gamma_{j k} \in \mathbb{C}$ ($j, k = 1, \dots, M$) are the elements of the decoherence matrix $\boldsymbol{\Gamma}\in \mathbb{C}^{M \times M}$, and the $\hat{F}_j$ correspond to the different decoherence channels.
It is worth noting that if $\boldsymbol{\Gamma}$ is diagonal, the master equation is already in diagonal Lindblad form.
For a system described by an effective master equation in a non-diagonal Lindblad form \eqref{eq:non-diagonal_lindblad_form}, given that $\boldsymbol{\Gamma}$ is a Hermitian positive semidefinite matrix which is a necessary and sufficient condition for the complete-positivity of the evolution, it is possible to bring the master equation into a diagonal Lindblad form \eqref{eq:diagonal_lindblad_form} by means of a unitary diagonalization of $\boldsymbol{\Gamma}$.
Nonetheless, after following the diagonalization procedure the obtained decoherence rates $\gamma_j$ are seldom simple functions of the original elements $\Gamma_{j k}$, which makes this method inadequate for the pursuit of analytical solutions.
Another plausible way to bring a master equation into a diagonal Lindblad form is to follow the diagonalization procedure described in Ref.~\cite{Hall2014}, where the master equation is brought into a ``canonical'' form, i.e., a master equation in diagonal Lindblad form where the decoherence operators correspond to an orthonormal basis set of traceless operators.
However, that does not make things any easier when it comes to solving the system dynamics.
It is thus significantly more convenient and efficient to work directly with the master equation in non-diagonal Lindblad form, and hence that is the direction we opt for in this paper.
\begin{figure}[t]
    \centering
    \includegraphics[width=\linewidth]{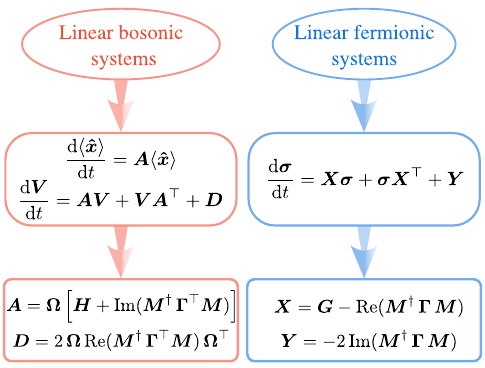}
    \caption{Summary of the main results of this work. 
    For linear bosonic systems, the dynamics is fully characterized by the evolution equations for the vector of means $\langle \boldsymbol{\x} \rangle$ and the symmetric covariance matrix $\boldsymbol{V}$, defined at the beginning of Sec.~\ref{sec:bosonic_linear_dynamics}. 
    The equations of motion shown correspond to Eqs.~\eqref{eq:vector_means_bosons} and \eqref{eq:lyapunov_bosons}, and the definitions of the drift and diffusion matrices $\boldsymbol{A}$ and $\boldsymbol{D}$ are also given in Eqs.~\eqref{eq:drift_matrix_bosons} and \eqref{eq:diffusion_matrix_bosons}.
    In the case of bosonic systems, $\boldsymbol{M}$ is a linear combination of quadrature operators given by Eq.~\eqref{eq:decoherence_operators}, while $\boldsymbol{H}$ and $\boldsymbol{\Omega}$ are given by Eqs.~\eqref{eq:symplectic_form} and \eqref{eq:hamiltonian_bosons}, respectively. 
    For linear fermionic systems, the dynamics of the system of interest is fully described by the evolution of the fermionic covariance matrix $\boldsymbol{\sigma}$, defined in the first part of  Sec.~\ref{sec:fermionic_linear_dynamics}. 
    The differential Lyapunov equation for $\boldsymbol{\sigma}$ corresponds to Eq.~\eqref{eq:lyapunov_fermions}, while the definitions of the drift and diffusion matrices $\boldsymbol{X}$ and $\boldsymbol{Y}$ are also given in Eqs.~\eqref{eq:drift_matrix_fermions} and \eqref{eq:diffusion_matrix_fermions}.
    In the case of fermionic systems, $\boldsymbol{M}$ is a linear combination of Majorana operators given by Eq.~\eqref{eq:decoherence_operators_fermions}, while $\boldsymbol{G}$ is defined in Eq.~\eqref{eq:hamiltonian_fermions}. 
    In both cases, $\boldsymbol{\Gamma}$ is the decoherence matrix defined in Eq.~\eqref{eq:non-diagonal_lindblad_form}. \label{fig:summary_results}}
\end{figure}
The main objective of this work is to present a set of equations that allow us to describe the dynamics of open quantum systems whose density matrix evolution is in a non-diagonal Lindblad form, as it arises naturally in quantum reservoir engineering.
In particular, we present evolution equations for the first and second moments of the canonical variables of bosonic and fermionic systems undergoing linear dynamics (see summary of the main results in Fig.~\ref{fig:summary_results}).
These equations are obtained from the parameters of the master equation in non-diagonal form and fully characterize the Gaussian dynamics of the system under study.
We emphasize that the method presented here provides an efficient way of undertaking the calculations regarding the study of linear engineered quantum systems.
Furthermore, it allows one to obtain analytical solutions for the system's steady state in a general and straightforward way, dramatically reducing the time required to do these typically tedious and cumbersome calculations.
Our main results are an extension of the method for constructing the dynamical equations for the first and second moments of the canonical variables from master equations in diagonal Lindblad form~\cite{Wiseman2005,Koga2012,Eisert2010,Bravyi2012}, and therefore, it is also our goal to draw attention to this type of approach, which we consider deserves to be more generally known.
This paper is organized as follows. In Sec.~\ref{sec:non-diagonal_master_equation}, we introduce the quantum master equation in non-diagonal Lindblad form and describe its relationship with the diagonal form.
In Sec.~\ref{sec:reservoir_engineering}, we consider a general open quantum system model to illustrate the multitude of scenarios in reservoir engineering in which master equations in non-diagonal Lindblad form may naturally appear.
In Sec.~\ref{sec:bosonic_linear_dynamics}, we describe the derivation of the dynamical equations for the first and second moments of the quadrature operators for bosonic linear systems described by non-diagonal Lindblad master equations, and show some covariance matrix techniques for the study of dissipative entanglement in this type of systems.
In Sec.~\ref{sec:fermionic_linear_dynamics}, following a procedure analogous to the one in the previous section, we discuss moment evolution equations and covariance matrix methods for linear fermionic systems described by master equations in non-diagonal Lindblad form.
In Sec~\ref{sec:ilustrative_examples}, we provide an illustrative example where we apply the proposed method to investigate dissipative entanglement between two non-interacting bosonic modes.
Conclusions are presented in Sec.~\ref{sec:conclusions}.
%


\section{Non-diagonal Lindblad master equations}
\label{sec:non-diagonal_master_equation}
In the theory of open quantum systems, the dynamics of the density matrix $\rho(t)$ describing the physical state of the system is given by a complete-positive trace-preserving (CPTP) dynamical map $\Phi(t,t_0)$, such that for all $t \geq t_0$ in the solution domain, $\rho(t)=\Phi(t,t_0) \rho(t_0)$~\cite{Breuer2007}.
The CPTP property of $\Phi(t,t_0)$ ensures that $\rho(t)$ is always positive semidefinite and trace one.
Nonetheless, in the Markov approximation, the system dynamics may be represented by a quantum dynamical semigroup, i.e., a single-parameter dynamical map $\Phi(t-t_0)$ satisfying the semigroup condition $\Phi(\tau_1 + \tau_2) = \Phi(\tau_1) \Phi(\tau_2)$ ($\tau_1, \tau_2 \geq 0$)~\cite{Alicki2007}. 
In this case, the dynamical map may be written as $\Phi(t-t_0)=e^{\L (t-t_0)}$, and the dynamics of the system density matrix will be governed by the quantum master equation 
\begin{align}\label{eq:master_equation}
    \dv{\rho}{t} = \L\, \rho,
\end{align}
where the Liouvillian $\L$ is the time-independent generator of the quantum dynamical semigroup~\cite{Walls2007}.
The most general characterization of the generator $\L$ was presented almost simultaneously in 1976 by Lindblad in Ref.~\cite{Lindblad1976}, and Gorini, Kossakowski, and Sudarshan in Ref.~\cite{Gorini1976}, which led to the GKLS master equation~\eqref{eq:diagonal_lindblad_form} (diagonal Lindblad form).
The master equation in diagonal Lindblad form may be written as 
\begin{align}
    &\dv{\rho}{t} = -i [\H, \rho] + \sum\limits_{j=1}^M \gamma_j\, \D[\Lk_j]\, \rho, \label{eq:diagonal_lindblad_form:2} 
\end{align}
where $\D[\Lk_j] \rho$ is the diagonal Lindblad dissipator given by
\begin{align}\label{eq:lindblad_dissipator}
    \mathcal{D}&[ \Lk_j ]\, \rho = \Lk_j\, \rho\, \Lk_j^{\dagger} - \frac{1}{2}\left( \Lk_j^{\dagger}\, \Lk_j\, \rho + \rho\, \Lk_j^{\dagger}\, \Lk_j \right).
\end{align}
Further, the complete positivity of the dynamical map is ensured if $\gamma_j \geq 0$ ($j=1, \dots, M$)~\cite{Breuer2015}.
However, as we mentioned before, in particular scenarios a more general type of dissipator appears.
This more general structure, which we refer to as the non-diagonal Lindblad form \eqref{eq:non-diagonal_lindblad_form}, includes the well-known diagonal Lindblad form (self-damping), but also enables the dissipative coupling between system operators (cross-damping).
The master equation in non-diagonal Lindblad form may be written as 
\begin{align}
    &\dv{\rho}{t} =  -i\, [\H, \rho] + \sum\limits_{j, k=1}^M \Gamma_{j k}\, \Ds[ \hat{F}_j, \hat{F}_k ]\, \rho, \label{eq:non-diagonal_lindblad:2} 
\end{align}
where $\Ds[ \hat{F}_j, \hat{F}_k ]\rho$ is a non-diagonal phase-dependent dissipator with the form
\begin{align}\label{eq:non-diagonal_dissipator}
    \Ds[ \hat{F}_j, \hat{F}_k ]\, \rho = \hat{F}_j\, \rho\, \hat{F}_k^\dagger - \frac{1}{2} \left( \hat{F}_k^\dagger\, \hat{F}_j\, \rho + \rho\, \hat{F}_k^\dagger\, \hat{F}_j \right),
\end{align}
and the $\Gamma_{j k}$ correspond to the elements of the decoherence matrix $\boldsymbol{\Gamma}\in \mathbb{C}^{M \times M}$.
The non-diagonal Lindblad form (\ref{eq:non-diagonal_lindblad_form},\ref{eq:non-diagonal_lindblad:2}) has also been called the Kossakowski-Lindblad form, and the matrix $\boldsymbol{\Gamma}$ has also been known as the Kossakowski matrix~\cite{Kossakowski1972}.
We avoid using that language since, in general, the operators $\hat{F}_j$ do not form an orthonormal basis for the space $\mathcal{B}(\mathcal{H})$ of bounded operators acting on the system's Hilbert $\mathcal{H}$.
Here, the complete positivity of the dynamical map is guaranteed if the decoherence matrix is a Hermitian positive semidefinite matrix ($\boldsymbol{\Gamma} \geq 0$).
Moreover, in accordance with Choi-Kraus' theorem on completely positive and trace-preserving maps, the number of decoherence operators $M$ for both diagonal and non-diagonal forms will be bounded from above by $d^2-1$, where $d$ is the dimension of the Hilbert space of the system~\cite{Manzano2020}. 
It is worth noting that cross-damping terms in master equations have been associated with the description of high-resolution spectroscopy of atomic systems \cite{Yost2014,Buchheit2016,Konovalov2020}, the Fano effect in cavity quantum electrodynamics \cite{Yamaguchi2021}, among others.
Moreover, master equations with cross-damping terms are ubiquitous in the theory of two-level systems coupled to a common bath, where they have been shown to be responsible for the dissipative entanglement between non-interacting qubits~\cite{Ficek2002}.
From Eqs.~\eqref{eq:lindblad_dissipator} and \eqref{eq:non-diagonal_dissipator}, it is clear that the diagonal dissipators correspond to the diagonal components of their non-diagonal counterparts ($\D[\Lk_j] \rho = \Ds[ \Lk_j, \Lk_j ] \rho$) and therefore, as mentioned above, the diagonal self-dissipation is included in the non-diagonal  description.
Furthermore, diagonal and non-diagonal forms are connected through the non-linearity of the diagonal dissipator, 
\begin{align}
    &\D[\alpha\, \Lk_j + \beta\, \Lk_k]\, \rho = |\alpha|^2\, \D[\Lk_j]\, \rho + |\beta|^2\, \D[\Lk_k]\, \rho \nonumber \\
    & \hspace{1.8cm} + \alpha \beta^*\, \D_s[\Lk_j,\Lk_k^\dagger]\, \rho + \alpha^* \beta\, \D_s[\Lk_k,\Lk_j^\dagger]\, \rho,
\end{align}
where $\alpha,\, \beta \in \mathbb{C}$.
This identity makes it evident that the decoherence associated with linear combinations of system operators will always lead to dissipative dynamics in a non-diagonal Lindblad form.
As mentioned in Sec.~\ref{sec:introduction}, since the decoherence matrix $\boldsymbol{\Gamma}$ is Hermitian and hence unitarily diagonalizable, it is always possible to bring a master equation in non-diagonal Lindblad form (\ref{eq:non-diagonal_lindblad_form},\ref{eq:non-diagonal_lindblad:2}) to a diagonal form (\ref{eq:diagonal_lindblad_form},\ref{eq:diagonal_lindblad_form:2}).
However, this procedure has its drawbacks.
On the one hand, if we want to find analytical expressions for the corresponding decoherence rates $\gamma_j$, one is limited by the possibility of obtaining closed-form expressions for the roots of the characteristic polynomial.
In fact, according to the Abel-Ruffini impossibility theorem, general polynomials of degree five or higher with arbitrary coefficients do not have simple algebraic solutions (i.e., in radicals).
This, of course, significantly limits the number $M$ of decoherence channels acting on the system, and therefore the size of the system itself, for which the $\gamma_j$ can be expressed as radicals of the $\Gamma_{jk}$; but even when possible, these expressions are usually quite involved.
On the other hand, if one opts for a numerical diagonalization of the decoherence matrix, since it is Hermitian, this can be done in $O(M^3)$ time~\cite{Banks2020}, being $M \times M$ the size of $\boldsymbol{\Gamma}$.
%
%
This corresponds to a computational complexity that grows as $O(d^6)$, where $d$ is the dimension of the Hilbert space of the system.
This is still polynomial time and is therefore computationally feasible; note, however, that any fine-tuning of the elements of the decoherence matrix involves a new numerical diagonalization, which is rather inefficient if other options are available.
Again, all this reinforces the argument we are making here that it is much more convenient to work with the master equations in non-diagonal Lindblad form directly.
For the derivation of the moment evolution equations below, it will be useful to calculate the adjoint of the Liouvillian $\L$ of a master equation in non-diagonal Lindblad form.
The adjoint of the generator of the quantum dynamical semigroup is defined according to 
\begin{align}\label{eq:definition_adjoint}
    \Tr [ \O (\L \rho) ] = \Tr [ (\L^\dagger \O) \rho ]
\end{align}
for an arbitrary operator $\O$, such that $\L^\dagger$ correspond to the adjoint of $\L$ with respect to the Hilbert-Schmidt inner product~\cite{Honda2010,Gyamfi2020}.
Further, if $\L$ is time-independent, the so-called adjoint master equation takes the form \cite{Breuer2007},
\begin{align}\label{eq:adjoint_master_equation}
    \dv{\O}{t} = \L^\dagger\, \O(t).
\end{align}
Thus, from the definition of the adjoint Liouvillian $\L^\dagger$ in Eq.~\eqref{eq:definition_adjoint} we find that the adjoint of the non-diagonal Lindblad master equation is given by
\begin{align}
    &\dv{\O}{t} =\, i [\H, \O] \nonumber \\
    & \hspace{0.8cm} + \sum\limits_{j,k=1}^M \Gamma_{j k} \left[ \hat{F}_k^\dagger \O \hat{F}_j - \frac{1}{2} \left( \O \hat{F}_k^\dagger \hat{F}_j + \hat{F}_k^\dagger \hat{F}_j \O \right) \right], \label{eq:adjoint_master_equation_gl} 
\end{align}
where $\H$, $\Gamma_{j k}$, and $\hat{F}_j$ correspond to the same entities defined above for Eq.~\eqref{eq:non-diagonal_lindblad_form}.
%


\section{Master equations in quantum reservoir engineering}
\label{sec:reservoir_engineering}
Quantum master equations in non-diagonal Lindblad form are ubiquitous in quantum reservoir engineering, where their appearance depends on particular properties of the reservoir state and the system-reservoir interaction.
These properties include, but are not limited to, squeezing of the reservoir's quantum state and the presence of explicit time-dependent terms in the system-reservoir interaction Hamiltonian.
Furthermore, a particular scenario where a non-diagonal description is essential is one in which distinct uncoupled quantum subsystems interact simultaneously with a common reservoir and, accordingly, the master equation is non-additive with respect to the subsystems involved.
Thus, if we consider a two-mode system, $\L \neq \L_1  + \L_2$ [with $\L$ as defined in Eq.~\eqref{eq:master_equation}] since an effective dissipative coupling between subsystems appears ($\L_{1 2}$), resulting in $\L = \L_1  + \L_2 + \L_{1 2}$; where $\L_i$ is the Liouvillian associated with the $i$-th subsystem ($i=1,2$), while $\L_{1 2}$ is a non-local Liouvillian.
Now, to illustrate why one should expect to obtain master equations in non-diagonal form in quantum reservoir engineering, we shall consider a system comprised of $N$ non-interacting modes (bosonic or fermionic) coupled to a common reservoir.
The total Hamiltonian for this open quantum system is,
\begin{align}\label{eq:hamiltonian}
    \H_{\mathrm{T}} = \H_{\mathrm{S}} + \H_{\mathrm{R}} + \H_{\mathrm{SR}},
\end{align}
where $\H_{\mathrm{S}}$ is the system Hamiltonian, $\H_{\mathrm{R}}$ describes the reservoir, and $\H_{\mathrm{SR}}$ is the system-reservoir interaction Hamiltonian.
Here, the system Hamiltonian takes the form ($\hbar=1$),
\begin{align}
   \H_{\mathrm{S}} = \sum_{j=1}^{N} \omega_j \a_j^\dagger \a_j, 
\end{align}
where $\omega_j$ is the frequency of the $j$-th system mode and $\a_j$ and $\a_j^\dagger$ are the annihilation and creation operators (bosonic or fermionic) associated with it, respectively.
In this model, the reservoir is comprised of infinite harmonic oscillators; however, to make the treatment as general as possible we will not fix the structure of the Hamiltonian $\H_{\mathrm{R}}$ but instead the reservoir will be characterized by the two-time correlation functions between the reservoir operators.
Moreover, we will consider an interaction Hamiltonian which in the Schrödinger picture is given by
\begin{align}\label{eq:sytem-reservoir_interaction}
    \H_{\mathrm{SR}}(t) = \sum_{j=1}^{N} \sum_{k=1}^\infty\, \b_k^\dagger\, \left[ g_{jk}^+(t)\, \a_j^\dagger + g_{jk}^-(t)\, \a_j \right] + \mathrm{H.c.},
\end{align}
where $\b_k$ and $\b_k^\dagger$ correspond to the annihilation and creation operators of the $k$-th reservoir harmonic oscillator, respectively.
Further, the $g_{jk}^\pm(t)$ are complex time-dependent couplings which may include phases associated with external coherent driving of the reservoir modes (usually after a linearization procedure), and may be written as
\begin{align}\label{eq:time-dependent_couplings}
    g_{jk}^\pm(t) = \sum_{l=0}^\infty c_{jk,l}^\pm e^{-i \Omega_{jk,l}^\pm t}\, ,
\end{align}
where $c_{jk,l}^\pm ,\, \Omega_{jk,l}^\pm \in \mathbb{R}$.
The structure of the interaction Hamiltonian in Eq.~\eqref{eq:sytem-reservoir_interaction} with the time-dependent couplings as defined in Eq.~\eqref{eq:time-dependent_couplings} is quite general for systems interacting linearly with a reservoir and, as we will see below, guarantees that the resulting master equation for the system's density matrix is in non-diagonal Lindblad form.
As is standard when deriving master equations describing reduced system dynamics, we move to an interaction picture with respect to $\H_0=\H_{\mathrm{S}} + \H_{\mathrm{R}}$, via $\H_{\mathrm{I}}= \hat{U}^{\dagger} \H_\mathrm{T} \hat{U} - i\, \hat{U}^{\dagger} \partial \hat{U}/\partial t$ with $\hat{U} = e^{-i \H_0 t}$. 
Thus, the total Hamiltonian in the interaction picture takes the form 
\begin{align}\label{eq:interaction_hamiltonian}
    \H_{\mathrm{I}}(t) = \sum_{k=1}^\infty \left[ \b_k^\dagger(t)\, \hat{A}_k(t) + \hat{A}_k^\dagger(t)\, \b_k(t)  \right],
\end{align}
where the $\hat{A}_k(t)$ are linear combinations of system operators given by
\begin{align}\label{eq:system_operators}
    \hat{A}_k(t) = \sum_{j=1}^{N} \left[ g_{jk}^+(t)\, e^{i \omega_j t}\, \a_j^\dagger + g_{jk}^-(t)\, e^{-i \omega_j t}\, \a_j \right],
\end{align}
with $\omega_j$ the frequency of the $j$-th system mode.
Further, $\b_k(t)= e^{i \H_{\mathrm{R}} t}\, \b_k\, e^{-i \H_{\mathrm{R}} t}$, such that the structure of $\b_k(t)$ will depend on the particular engineering of the reservoir.
In the Born-Markov approximations, the evolution of the reduced system density matrix will be given by~\cite{Carmichael1999,Walls2007},
\begin{align}\label{eq:born-markov}
    \dv{\rho(t)}{t}=-\int_{0}^{+\infty} \dd s\, \Tr_{\R} \left[ \H_\mathrm{I}(t),\left[\H_\mathrm{I}(t-s), \rho(t) \otimes\rho_{\R}\right] \right] ,
\end{align}
where $\rho(t) = \Tr_{\R} [\rho_\tot]$, with $\rho_\tot$ the total density matrix (system plus reservoir).
It is important to keep in mind that Eq.~\eqref{eq:born-markov} is valid if the system is weakly coupled to the reservoir (Born approximation), and the reservoir correlations decay rapidly compared with the relevant time scales of the system (Markov approximation).
From Eq.~\eqref{eq:born-markov}, we have
\begin{align}\label{eq:born-markov_2}
    \dv{\rho(t)}{t} = \int_{0}^{+\infty}& \dd s\, \Tr_{\R} \left[ \H_\mathrm{I}(t-s)\, \rho(t) \otimes\rho_{\R}\, \H_\mathrm{I}(t) \right. \nonumber \\
    & \left. -  \H_\mathrm{I}(t) \H_\mathrm{I}(t-s)\, \rho(t) \otimes\rho_{\R} \right] + \mathrm{H.c.},
\end{align}
where
\begin{subequations}\label{eq:trace}
\begin{align}
    &\Tr_{\R} \left[ \H_\mathrm{I}(t-s)\, \rho(t) \otimes\rho_{\R}\, \H_\mathrm{I}(t) \right] \nonumber \\
    & = \sum_{k,k'=1}^\infty   \left[\, \langle\, \b_k(t)\, \b_{k'}^\dagger(t-s)  \rangle\,  \hat{A}_{k'}(t-s)\, \rho(t)\, \hat{A}_k^\dagger(t)  \right. \nonumber \\
    & \hspace{1.15cm} + \langle\, \b_k^\dagger(t)\, \b_{k'}(t-s)  \rangle\,  \hat{A}_{k'}^\dagger(t-s)\, \rho(t)\, \hat{A}_k(t) \nonumber \\
    & \hspace{1.15cm} + \langle\, \b_k(t)\, \b_{k'}(t-s)  \rangle\,   \hat{A}_{k'}^\dagger(t-s)\, \rho(t)\, \hat{A}_k^\dagger(t) \nonumber \\
    & \left. \hspace{1.15cm} + \langle\, \b_k^\dagger(t)\, \b_{k'}^\dagger(t-s)  \rangle\,  \hat{A}_{k'}(t-s)\, \rho(t)\, \hat{A}_k(t)
    \, \right],
\end{align}
\begin{align}
    &\Tr_{\R} \left[ \H_\mathrm{I}(t) \H_\mathrm{I}(t-s)\, \rho(t) \otimes\rho_{\R} \right] \nonumber \\
    & = \sum_{k,k'=1}^\infty   \left[\, \langle\, \b_k(t)\, \b_{k'}^\dagger(t-s)  \rangle\,  \hat{A}_k^\dagger(t) \hat{A}_{k'}(t-s)\, \rho(t) \right. \nonumber \\
    & \hspace{1.15cm} + \langle\, \b_k^\dagger(t)\, \b_{k'}(t-s)  \rangle\,  \hat{A}_k(t) \hat{A}_{k'}^\dagger(t-s)\, \rho(t) \nonumber \\
    & \hspace{1.15cm} + \langle\, \b_k(t)\, \b_{k'}(t-s)  \rangle\,  \hat{A}_k^\dagger(t) \hat{A}_{k'}^\dagger(t-s)\, \rho(t) \nonumber \\
    & \left. \hspace{1.15cm} + \langle\, \b_k^\dagger(t)\, \b_{k'}^\dagger(t-s) \rangle\,  \hat{A}_k(t) \hat{A}_{k'}(t-s)\, \rho(t)
    \, \right].
\end{align}
\end{subequations}
Here we will assume that the correlations between the different reservoir modes decay fast enough so that the correlation functions in Eqs.~\eqref{eq:trace} are proportional to $\delta_{k k'}$.
Further, we shall consider that $\rho_\R$ is a stationary state ($[\H_\R, \rho_\R] =  0$) and, accordingly, the reservoir correlation functions are time-homogeneous.
Thus, we have
\begin{subequations}\label{eq:correlation_functions}
\begin{align}
    \langle \b_k(t)\b^\dagger_{k'}(t-s) \rangle = \langle \b_k(s)\b^\dagger_k(0) \rangle\, \delta_{k k'}, \label{eq:correlation_functions:a} \\
    \langle \b^\dagger_k(t)\b_{k'}(t-s) \rangle = \langle \b^\dagger_k(s)\b_k(0) \rangle\, \delta_{k k'}, \label{eq:correlation_functions:b} \\
    \langle \b^\dagger_k(t)\b^\dagger_{k'}(t-s) \rangle = \langle \b^\dagger_k(s)\b^\dagger_k(0) \rangle\, \delta_{k k'}, \\
    \langle \b_k(t)\b_{k'}(t-s) \rangle = \langle \b_k(s)\b_k(0) \rangle\, \delta_{k k'}.
\end{align}
\end{subequations}
We remark that there are some specially engineered reservoirs for which the reservoir state is nonstationary and the correlation functions are not time-homogeneous, as for instance when the reservoir is in a squeezed state.
For simplicity we will assume that the phase-dependent correlation functions are null, $\langle \b_k(s)\b_k(0) \rangle = \langle \b_k^\dagger(s)\b_k^\dagger(0) \rangle = 0$, which is true in the vast majority of practical examples where $\rho_\R$ is stationary.
Now, taking into account Eqs.~\eqref{eq:correlation_functions:a} and \eqref{eq:correlation_functions:b}, we must explicitly calculate each of the terms in Eqs.~\eqref{eq:trace}, and apply the secular approximation to them.
To show the general procedure to follow, let us consider the product $\hat{A}_k^\dagger(t) \hat{A}_k(t-s)$ which, according to Eqs.~\eqref{eq:time-dependent_couplings} and \eqref{eq:system_operators}, takes the form
\begin{align}\label{eq:secular_approximation}
    &\hat{A}_k^\dagger(t) \hat{A}_k(t-s) = \sum_{j,j'=1}^{N}  \sum_{l,l'=0}^{\infty}  \nonumber \\
    & \hspace{0.2cm} \left[\,  c_{jk,l}^-\, c_{j'k,l'}^-\, e^{i (\Omega_{jk,l}^- + \omega_j) t}\, e^{-i (\Omega_{j'k,l'}^- + \omega_{j'}) (t-s)}\, \a_j^\dagger \a_{j'} \right. \nonumber \\
    & + c_{jk,l}^+\, c_{j'k,l'}^+\, e^{i (\Omega_{jk,l}^+ - \omega_j) t}\, e^{-i (\Omega_{j'k,l'}^+ -  \omega_{j'}) (t-s)}\, \a_j \a_{j'}^\dagger \nonumber \\
    & +  c_{jk,l}^+\, c_{j'k,l'}^-\, e^{i (\Omega_{jk,l}^+ - \omega_j) t}\, e^{-i (\Omega_{j'k,l'}^- + \omega_{j'}) (t-s)}\, \a_j \a_{j'} \nonumber \\
    & \left. + c_{jk,l}^-\, c_{j'k,l'}^+\, e^{i (\Omega_{jk,l}^- + \omega_j) t}\, e^{-i (\Omega_{j'k,l'}^+ -  \omega_{j'}) (t-s)}\, \a_j^\dagger \a_{j'}^\dagger \, \right].
\end{align}
We neglect here the terms that oscillate rapidly with respect to the time variable $t$ (secular approximation), so that
\begin{align}
    &\hat{A}_k^\dagger(t) \hat{A}_k(t-s) \approx \sum_{j,j'=1}^{N} \left[\, G^{(1,1)}_{j j',k}(s)\, \a_j^\dagger \a_{j'} \right. \nonumber \\
    & \left. + G^{(2,1)}_{j j',k}(s)\, \a_j \a_{j'} + G^{(3,1)}_{j j',k}(s)\, \a_j^\dagger \a_{j'}^\dagger + G^{(4,1)}_{j j',k}(s)\, \a_j \a_{j'}^\dagger \, \right];
\end{align}
with
\begin{subequations}\label{eq:g_1}
\begin{align}
    &G^{(1,1)}_{j j',k}(s) = \sum_{l,l'=0 }^{\infty}  c_{jk,l}^-\, c_{j'k,l'}^-\, e^{i (\Omega_{j'k,l'}^- +  \omega_{j'}) s}\, ,\\[-5pt]
    & \hspace{0.95cm} {\scriptstyle (\Omega_{j'k,l'}^- +  \omega_{j'} = \Omega_{j k,l}^- + \omega_j)} \nonumber \\[5pt]
    &G^{(2,1)}_{j j',k}(s) = \sum_{l,l'=0 }^{\infty}  c_{jk,l}^+\, c_{j'k,l'}^-\, e^{i (\Omega_{j'k,l'}^- +  \omega_{j'}) s}\, , \\[-5pt]
    & \hspace{0.95cm} {\scriptstyle (\Omega_{j'k,l'}^- +   \omega_{j'} = \Omega_{jk,l}^+ - \omega_j)} \nonumber \\[5pt]
    &G^{(3,1)}_{j j',k}(s) = \sum_{l,l'=0 }^{\infty}  c_{jk,l}^-\, c_{j'k,l'}^+\, e^{i (\Omega_{j'k,l'}^+ -  \omega_{j'}) s}\, , \\[-5pt]
    & \hspace{0.95cm} {\scriptstyle (\Omega_{j'k,l'}^+ -  \omega_{j'} = \Omega_{jk,l}^- + \omega_j)} \nonumber \\[5pt]
    &G^{(4,1)}_{j j',k}(s) = \sum_{l,l'=0 }^{\infty}  c_{jk,l}^+\, c_{j'k,l'}^+\, e^{i (\Omega_{j'k,l'}^+ -  \omega_{j'}) s}\, . \\[-5pt]
    & \hspace{0.95cm} {\scriptstyle (\Omega_{j'k,l'}^+ -  \omega_{j'} = \Omega_{jk,l}^+ - \omega_j)} \nonumber 
\end{align}
\end{subequations}
The resonance condition in parentheses below the summation symbol must be satisfied for all elements in the sum.
The index $n$ in the superscript $(m,n)$ of the $G_{j k, l}^{(m,n)}(s)$ refers to the correlation function that accompanies these terms, $n=1$ for $\langle \b_l(s)\b_l^\dagger(0) \rangle$ and $n=2$ for $\langle \b_l^\dagger(s)\b_l(0) \rangle$.
When making the secular approximation above, we neglected terms with an explicit time-dependence of the form $e^{i (\epsilon-\epsilon') t}$, for $\epsilon \neq \epsilon'$.
The validity of this approximation is based on the assumption that the intrinsic time-scale $\tau_{\S} = ( \mathrm{min}_{\epsilon \neq \epsilon'} |\epsilon-\epsilon'|)^{-1}$ is much shorter than the relaxation time $\tau_{\R}$ over which the system changes appreciably and, accordingly, the counter-rotating terms will have a very small effect on the overall dynamics of the system density matrix~\cite{Breuer2007,Yamaguchi2017}.
Hence, the suppression of the counter-rotating terms in Eq.~\eqref{eq:secular_approximation}, and thereby the secular approximation, corresponds to a coarse-grained time average.
It is important to note that if the condition $\tau_{\S} \ll \tau_{\R}$ is not satisfied, making a secular approximation may yield unphysical results~\cite{Wichterich2007,Esposito2009}.
In that case we would have slowly oscillating terms that cannot be eliminated, and a partial secular approximation would be required~\cite{Jeske2015,Cresser2017,Farina2019,Cattaneo2019}.
Following an analogous procedure for $\hat{A}_k(t) \hat{A}_k^\dagger(t-s)$, we obtain
\begin{align}
    &\hat{A}_k(t) \hat{A}_k^\dagger(t-s) \approx \sum_{j,j'=1}^{N} \left[\, G^{(1,2)}_{j j',k}(s)\, \a_j^\dagger \a_{j'} \right. \nonumber \\
    & \left. + G^{(2,2)}_{j j',k}(s)\, \a_j \a_{j'} + G^{(3,2)}_{j j',k}(s)\, \a_j^\dagger \a_{j'}^\dagger + G^{(4,2)}_{j j',k}(s)\, \a_j \a_{j'}^\dagger \, \right];
\end{align}
where the $G_{j k, l}^{(m,2)}(s)$ are related to the $G_{j k, l}^{(m,1)}(s)$ through the following relationships,
\begin{subequations}\label{eq:g_relationships}
\begin{align}
    G_{j k, l}^{(1,2)}(s) &= G_{j k, l}^{(4,1) *}(s), \\
    G_{j k, l}^{(2,2)}(s) &= G_{j k, l}^{(3,1) *}(s), \\
    G_{j k, l}^{(3,2)}(s) &= G_{j k, l}^{(2,1) *}(s), \\
    G_{j k, l}^{(4,2)}(s) &= G_{j k, l}^{(1,1) *}(s). 
\end{align}
\end{subequations}
Thus, we find that the master equation \eqref{eq:born-markov} for the reduced system dynamics takes the form 
\begin{align}\label{eq:master_equation_1}
    \dv{\rho(t)}{t} = \sum_{j,k = 1}^{N} & \,\, \left\{ \, \gamma_{j k}^{(1)}\, \left[  \a_j\, \rho(t)\, \a_k^\dagger - \a_k^\dagger \a_j\, \rho(t) \right] \right. \nonumber \\
    & + \gamma_{j k}^{(2)}\, \left[  \a_j\, \rho(t)\, \a_k - \a_k \a_j\, \rho(t) \right] \nonumber \\
    &  + \gamma_{j k}^{(3)}\, \left[  \a_j^\dagger\, \rho(t)\, \a_k^\dagger - \a_k^\dagger \a_j^\dagger\, \rho(t) \right] \nonumber \\
    &  \left. +  \gamma_{j k}^{(4)}\, \left[  \a_j^\dagger\, \rho(t)\, \a_k - \a_k \a_j^\dagger\, \rho(t) \right]\, \right\} + \mathrm{H.c.},
\end{align}
where 
\begin{align}
    &\gamma_{j k}^{(m)} = \int_0^\infty \dd{s}\, \sum_{l=1}^\infty \, \left[\,
    G_{j k, l}^{(m,1)}(s)\, \langle \b_l(s)\b_l^\dagger(0) \rangle \right. \nonumber \\
    & \left. \hspace{3.5cm} + G_{j k, l}^{(m,2)}(s)\, \langle \b_l^\dagger(s)\b_l(0) \rangle \, \right].
\end{align}
As shown above, the $G_{j k, l}^{(m,n)}(s)$ come from considering the secular approximation, so that the terms oscillating rapidly with respect to the time-variable $t$ were neglected.
Furthermore, the engineering of the quantum reservoir will modify the structure of the correlation functions $\langle \b_l(s)\b_l^\dagger(0) \rangle$ and $\langle \b_l^\dagger(s)\b_l(0) \rangle$, so that we will be able to tailor the dissipative dynamics of the system.
From Eqs.~\eqref{eq:g_1} and \eqref{eq:g_relationships}, we can show that $G_{j k, l}^{(1,n)}(s) = G_{k j, l}^{(1,n)}(s)$, $G_{j k, l}^{(2,n)}(s) = G_{k j, l}^{(3,n)}(s)$, and $G_{j k, l}^{(4,n)}(s) = G_{k j, l}^{(4,n)}(s)$, which implies that,
\begin{subequations}\label{eq:gamma_symmetry}
\begin{align}
    &\gamma_{jk}^{(1)} = \gamma_{kj}^{(1)}, \\
    &\gamma_{jk}^{(2)} = \gamma_{kj}^{(3)}, \\
    &\gamma_{jk}^{(4)} = \gamma_{kj}^{(4)}.
\end{align}
\end{subequations}
Taking this into account, we now make $\gamma_{j k}^{(m)} = \Gamma_{j k}^{(m)}/2 + i \Upsilon_{j k}^{(m)}$, such that Eq.~\eqref{eq:master_equation_1} may be written as
\begin{align}\label{eq:general_microscopic_master_equation}
    \dv{\rho(t)}{t} = -i [\H_\mathrm{LS}, \rho(t)] +  \sum_{j,k=1}^{N} \left\{\, \Gamma_{j k}^{(1)}\, \Ds[\a_j, \a_k]\, \rho(t) \right. \nonumber \\
    + \Gamma_{j k}^{(2)}\, \Ds[\a_j, \a_k^\dagger]\, \rho(t) + \Gamma_{j k}^{(3)}\, \Ds[\a_j^\dagger, \a_k]\, \rho(t) \nonumber \\
    \left. + \Gamma_{j k}^{(4)}\, \Ds[\a_j^\dagger, \a_k^\dagger]\, \rho(t)\,  \right\},
\end{align}
where
\begin{align}
    &\H_\mathrm{L S} =   \sum_{j,k=1}^{N} \, \left[ \, \Upsilon_{j k}^{(1)} \a_j^\dagger \a_k + \Upsilon_{j k}^{(2)} \a_j^\dagger \a_k^\dagger \right. \nonumber \\ 
    & \left. \hspace{3.5cm} + \Upsilon_{j k}^{(3)} \a_j \a_k + \Upsilon_{j k}^{(4)} \a_j \a_k^\dagger \right]
\end{align}
is a Lamb shift Hamiltonian, whereas $\Ds[ \hat{F}_j, \hat{F}_k ]\rho$ is the non-diagonal Lindblad dissipator defined in Eq.~\eqref{eq:non-diagonal_dissipator}, which enables the dissipative coupling of the different modes of the system through the reservoir.
We may now organize the system creation and annihilation operators in a vector of decoherence operators, $\boldsymbol{\hat{F}} = (\a_1, \dots, \a_N, \a_1^\dagger, \dots, \a_N^\dagger)^\top$, and the associated rates in a decoherence matrix, 
\begin{align}
\boldsymbol{\Gamma} = \left(\begin{array}{cc}
    \boldsymbol{\Gamma}^{(1)} & \boldsymbol{\Gamma}^{(2)} \\
    \boldsymbol{\Gamma}^{(3)} & \boldsymbol{\Gamma}^{(4)}
    \end{array}\right),
\end{align}
where the elements of each $N \times N$ block are given by the $\Gamma_{j k}^{(l)}$ rates.
From Eqs.~\eqref{eq:gamma_symmetry} it follows that $\boldsymbol{\Gamma}^{(1)} = \boldsymbol{\Gamma}^{(1)\, \top}$, $\boldsymbol{\Gamma}^{(2)} = \boldsymbol{\Gamma}^{(3)\, \top}$, and $\boldsymbol{\Gamma}^{(4)} = \boldsymbol{\Gamma}^{(4)\, \top}$, which yields that $\boldsymbol{\Gamma}$ is symmetric ($\boldsymbol{\Gamma} = \boldsymbol{\Gamma}^{\top}$).
Moreover, as a consequence of Bochner's theorem,  any physical correlation function must be positive semidefinite, which in turn implies that the decoherence matrix $\boldsymbol{\Gamma}$  will always be positive semidefinite as well~\cite{Yaglom1987}.
Therefore, in the Schrödinger picture Eq.~\eqref{eq:general_microscopic_master_equation} takes the form
\begin{align}\label{eq:master_equation_2}
    &\dv{\rho(t)}{t} =  -i [\H, \rho(t)] + \sum\limits_{j, k=1}^{2 N} \Gamma_{j k}\, \Ds[ \hat{F}_j, \hat{F}_k ]\, \rho(t),
\end{align}
where $\H$ is a system Hamiltonian that includes the Lamb shift correction.
Clearly, the master equation~\eqref{eq:master_equation_2} is in the non-diagonal  Lindblad form~(\ref{eq:non-diagonal_lindblad_form},\ref{eq:non-diagonal_lindblad:2}).
However, it is worth mentioning that in this case the appearance of cross-damping terms in the master equation (non-diagonal decoherence matrix) is subject to at least one of the $G_{j k, l}^{(2,n)}(s) = G_{k j, l}^{(3,n)}(s)$ being different from zero, or to the existence of some non-zero $G_{j k, l}^{(1,n)}(s)$ or $G_{j k, l}^{(4,n)}(s)$ for which $j\neq k$.
Further, the existence of any $G_{j k, l}^{(m,n)}(s) \neq 0$ depends on satisfying at least once the resonance condition that defines it.
We remark that this type of conditions are easily met in quantum reservoir engineering.
In the next two sections, we present a theoretical framework in which the dynamics of linear systems described by master equations in non-diagonal Lindblad form can be solved exactly.
%


\section{Linear dynamics of bosonic systems}
\label{sec:bosonic_linear_dynamics}
For an $N$-mode continuous-variable (CV) system, described by the creation and annihilation operators $\a^\dagger_j$ and $\a_j$ satisfying the canonical commutation relation $[\a_{j}, \a_k^\dagger]=\delta_{j k}$ ($j, k = 1, \dots N$), the full information on the quantum state of the system can be reconstructed from the statistics of the canonical quadrature operators involved,  $\hat{q}_j = (\a^\dagger_j + \a_j)/\sqrt{2}$ and $\hat{p}_j = i (\a^\dagger_j - \a_j)/\sqrt{2}$ with $[\hat{q}_j,\hat{p}_k]= i \delta_{j k}$~\cite{Smithey1993}.
Furthermore, Gaussian states are fully characterized by the first and second moments of the quadrature operators of the system, since they obey Wick's theorem, which allows one to calculate the expectation values of arbitrary products of operators from expectation values of products of pairs~\cite{Wick1950, Hackl2021}.
Thus, if we group the canonical quadratures in a vector of quadrature operators $\boldsymbol{\x} = \left(\hat{q}_{1}, \hat{p}_{1}, \dots, \hat{q}_{N}, \hat{p}_{N} \right)^\top$, all the relevant properties of the system will be described by the vector of means $\langle \boldsymbol{\x} \rangle = \Tr[\boldsymbol{\x} \rho ]$ and the symmetrically-ordered covariance matrix $\boldsymbol{V}\in\mathbb{R}^{2N \times 2N}$, whose elements are defined as $V_{j k} = \langle \{ \Delta \x_j, \Delta \x_k \} \rangle = \langle \{ \x_j, \x_k \} \rangle - 2 \langle \x_j \rangle \langle \x_k \rangle$. 
This phase-space representation is known as the covariance matrix picture or the symplectic picture, and enables the exact treatment of quadratic systems.
It is then worthwhile to develop methods that allow us to determine the dynamics of the state and observables of an open quantum system from the vector of means and the covariance matrix.
These methods, often called covariance matrix methods, require techniques for finding the evolution equations of the statistical moments of the canonical quadratures without resorting to the explicit solution of the density matrix dynamics.
In the past, these techniques have been presented in the context of Gaussian dynamics described by master equations in diagonal Lindblad form~\cite{Wiseman2005,Koga2012}.
Here, we make an extension to systems described by master equations in non-diagonal Lindblad form, where it is in general inefficient to make a transformation to a diagonal form.
An $N$-mode bosonic CV system is defined on the Hilbert space $\mathcal{H}= \bigotimes_{j=1}^N \mathcal{H}_{j}$, where each $\mathcal{H}_{j}$ is the infinite-dimensional Hilbert space of a harmonic oscillator.
In this space, as mentioned above, the state of the system may be described by the statistical moments of the quadrature operators arranged in the vector $\boldsymbol{\x}$, where the canonical commutation relation may be written as
\begin{align}\label{eq:commutation_relation}
    \boldsymbol{\x} \boldsymbol{\x}^\top- \left( \boldsymbol{\x} \boldsymbol{\x}^\top \right)^\top = i \boldsymbol{\Omega},
\end{align}
with $\boldsymbol{\Omega}\in \mathbb{R}^{2N\times 2N}$ the $N$-mode symplectic form given by
\begin{align}\label{eq:symplectic_form}
    \boldsymbol{\Omega} = \bigoplus_{j=1}^N \boldsymbol{\omega}, \quad \boldsymbol{\omega} = \left(\begin{array}{cc}
    0 & 1 \\
    -1 & 0
    \end{array}\right).
\end{align}
In index notation, Eq.~\eqref{eq:commutation_relation} takes the form $\left[\x_{j}, \x_{k}\right]= i \Omega_{j k}$.
Moreover, a given covariance matrix will be a valid descriptor of a physical quantum state on $\mathcal{H}$, if and only if the Robertson–Schrödinger uncertainty relation $\boldsymbol{V} + i \boldsymbol{\Omega} \geq 0$
is satisfied~\cite{Olivares2012,Adesso2014,Serafini2017}.
In other words, the Hermitian matrix $\boldsymbol{V} + i \boldsymbol{\Omega}$ has to be positive semidefinite, and since $\boldsymbol{\Omega}$ is antisymmetric it follows that $\boldsymbol{V}$ must be positive semidefinite, $\boldsymbol{V} \geq 0$.
Interestingly enough, the above uncertainty inequality is the consequence of considering the canonical commutation relation \eqref{eq:commutation_relation} and the positivity of the quantum state $\rho$ and, therefore, the uncertainty relation is a necessary condition for the system density matrix to be positive semidefinite~\cite{Serafini2017}.
Furthermore, it is worth noting that the purity of a bosonic Gaussian state is given by the formula $\mu = \Tr [ \rho^2 ] = 1/\sqrt{\mathrm{det}(\boldsymbol{V})}$~\cite{Adesso2014}, such that a pure Gaussian state satisfies $\mathrm{det}(\boldsymbol{V}) = 1$.
Alternatively, for a pure bosonic Gaussian state, all eigenvalues of the matrix $(\boldsymbol{\Omega} \boldsymbol{V})^2$ must be equal to unity~\cite{Eisert2010}.
Now, let us consider an $N$-mode bosonic open quantum system that evolves according to the master equation in non-diagonal Lindblad form \eqref{eq:non-diagonal_lindblad_form}, where the Hamiltonian is quadratic and the decoherence operators are linear with respect to the canonical quadrature operators.
Systems satisfying these characteristics are called linear, since the equations of motion for the canonical operators are linear, and any measurement corresponds to linear combinations of these.
Linear dynamics ensures that the equations of motion for the vector of means and the covariance matrix are closed and hence sufficient to describe the state of initially Gaussian systems at any point in time~\cite{Wiseman2009}.
Thus, linear dynamics preserves the Gaussian character of physical states and, therefore, is of utmost importance in quantum information theory with continuous variables~\cite{Adesso2014}.
Further, through dissipative linear dynamics non-Gaussian initial states can be transformed into Gaussian states~\cite{Jacobs2014}.
The Hamiltonian of the linear system under consideration has the form
\begin{align}\label{eq:hamiltonian_bosons}
    \H = \frac{1}{2} \boldsymbol{\x}^\top \boldsymbol{H}\, \boldsymbol{\x},
\end{align}
which may be written in index notation as $\H = \frac{1}{2} \sum_{j,k = 1}^N H_{j k}\, \x_j \x_k$, where $\boldsymbol{H}\in \mathbb{R}^{2N \times 2N}$ is a symmetric matrix known as the Hamiltonian matrix.
Moreover, the elements of the vector of decoherence operators $\mathbf{\hat{F}}$ may be written as linear combinations of quadrature operators, so that we have
\begin{align}\label{eq:decoherence_operators}
    \hat{F}_j = \sum_{k=1}^{2N} M_{j k}\, \x_k, 
\end{align}
with $M_{j k}$ the elements of the transformation matrix $\boldsymbol{M}\in \mathbb{C}^{M \times 2N}$.
To calculate the dynamics of the covariance matrix and the vector of means of a linear system, the standard approach consists of taking into account that $\langle \dot{\O} \rangle = \Tr [ \O\, \L \rho(t) ]$  for the particular Liouvillian $\L$ of the model under study, and consider both $\O = \x_j$ and $\O=\x_j \x_k$ to find a closed system of equations for $\langle \boldsymbol{\x} \rangle$ and $\boldsymbol{V}$ (see, e.g., Refs.~\cite{Malouf2019,Malouf2020}).
Such an approach is often cumbersome and time-consuming, and completely lacks generality.
However, general formulas exist (see Refs.~\cite{Wiseman2005,Koga2012}), but they only consider systems described by master equations in diagonal Lindblad form.
Thus, we find below a set of formulas that relate the equations of motion for $\langle \boldsymbol{\x} \rangle$ and $\boldsymbol{V}$ to the parameters of a master equation in non-diagonal Lindblad form describing the dynamics of a linear system.
To determine the equations of motion for the vector of means and the covariance matrix, we shall consider that $\langle \dot{\O} \rangle = \Tr [ \dot{\O}(t)\, \rho ] = \langle \L^\dagger \O(t) \rangle$ with $\L^\dagger$ the adjoint of the Lindblad generator defined in Eqs.~(\ref{eq:adjoint_master_equation},\ref{eq:adjoint_master_equation_gl}). 
Hence, the time derivative of the vector of means will take the form
\begin{align}\label{eq:mean_adjoint_bosons}
    \dv{\langle \boldsymbol{\x} \rangle}{t} =  \langle \L^\dagger \boldsymbol{\x} \rangle.
\end{align}
This leads to the general expression (see the \hyperref[app:structure_adjoint_equation]{Appendix\ref{app:structure_adjoint_equation}} for more details),
\begin{align}\label{eq:vector_means_bosons}
    \dv{\langle \boldsymbol{\x} \rangle}{t} = \boldsymbol{A} \langle \boldsymbol{\x} \rangle,
\end{align}
where $\boldsymbol{A}\in \mathbb{C}^{2N \times 2N}$ is given by
\begin{align}\label{eq:drift_matrix_bosons_0}
    \boldsymbol{A} = \boldsymbol{\Omega} \boldsymbol{H} + \frac{i}{2} \boldsymbol{\Omega}\, \left[  (\boldsymbol{M}^\dagger\, \boldsymbol{\Gamma}^*\, \boldsymbol{M})^* - \boldsymbol{M}^\dagger\, \boldsymbol{\Gamma}^\top \boldsymbol{M} \right],
\end{align}
with $\boldsymbol{\Gamma}$, $\boldsymbol{\Omega}$, $\boldsymbol{H}$, and $\boldsymbol{M}$ as defined in Eqs.~\eqref{eq:non-diagonal_lindblad_form}, \eqref{eq:symplectic_form}, \eqref{eq:hamiltonian_bosons}, and \eqref{eq:decoherence_operators}, respectively.
Analogously, the evolution of the elements of the covariance matrix will be given by
\begin{align}\label{eq:elements_cm}
    \dv{ V_{j k} }{t} =  \langle \L^\dagger (\x_j \x_k + \x_k \x_j) \rangle - 2 \dv{}{t} \left( \langle \x_j \rangle \langle \x_k \rangle \right),
\end{align}
where the second term on the right-hand side can be determined from Eq.~\eqref{eq:vector_means_bosons}.
From Eq.~\eqref{eq:elements_cm} we find the differential Lyapunov equation
\begin{align}\label{eq:lyapunov_bosons}
    \dv{\boldsymbol{V}}{t} = \boldsymbol{A} \boldsymbol{V} + \boldsymbol{V} \boldsymbol{A}^\top + \boldsymbol{D},
\end{align}
where the drift matrix $\boldsymbol{A}$ is given by Eq.~\eqref{eq:drift_matrix_bosons_0} and the diffusion matrix $\boldsymbol{D}$ is
\begin{align}
    \boldsymbol{D} = \boldsymbol{\Omega}\, \left[ (\boldsymbol{M}^\dagger\, \boldsymbol{\Gamma}^*\, \boldsymbol{M})^* + \boldsymbol{M}^\dagger\, \boldsymbol{\Gamma}^\top \boldsymbol{M}  \right]\, \boldsymbol{\Omega}^\top.
\end{align}
Finally, given that $\boldsymbol{\Gamma}$ is Hermitian, which is also a necessary condition so that $\boldsymbol{V}(t)$ is real, we obtain
\begin{subequations}\label{eq:drift_diffusion_bosons}
\begin{align}
    \boldsymbol{A} &= \boldsymbol{\Omega} \left[ \boldsymbol{H} +  \Im ( \boldsymbol{M}^\dagger\, \boldsymbol{\Gamma}^\top \boldsymbol{M}) \right], \label{eq:drift_matrix_bosons} \\
    \boldsymbol{D} &= 2\, \boldsymbol{\Omega}\, \Re (  \boldsymbol{M}^\dagger\, \boldsymbol{\Gamma}^\top \boldsymbol{M} )\, \boldsymbol{\Omega}^\top. \label{eq:diffusion_matrix_bosons}
\end{align}
\end{subequations}
The equations of motion \eqref{eq:vector_means_bosons} and \eqref{eq:lyapunov_bosons}, together with the definitions for $\boldsymbol{A}$ and $\boldsymbol{D}$ in Eqs.~\eqref{eq:drift_matrix_bosons} and \eqref{eq:diffusion_matrix_bosons}, represent the first part of the main results of this paper, which correspond to the dynamical equations for the vector of means and the covariance matrix of a bosonic linear system whose density matrix evolution is described by a master equation in non-diagonal Lindblad form.
It is important to note that if the master equation describing the dynamics of the linear system is in diagonal Lindblad form~\eqref{eq:diagonal_lindblad_form}, $\boldsymbol{\Gamma} = \mathrm{diag}( \gamma_1, \dots, \gamma_M) \in \mathbb{R}^{M\times M}$, and we may include the decoherence parameters in the decoherence operators so that we introduce the matrix $\boldsymbol{C} = \sqrt{\boldsymbol{\Gamma}} \boldsymbol{M}\in \mathbb{C}^{M \times 2N}$.
Thus, the drift and diffusion matrices in Eqs.~\eqref{eq:vector_means_bosons} and \eqref{eq:lyapunov_bosons} would take the known form $\boldsymbol{A} = \boldsymbol{\Omega} [\boldsymbol{H} + \Im (\boldsymbol{C}^\dagger \boldsymbol{C})]$ and $\boldsymbol{D} = 2\, \boldsymbol{\Omega} \Re (\boldsymbol{C}^\dagger \boldsymbol{C}) \boldsymbol{\Omega}^\top$~\cite{Wiseman2005,Koga2012}.
The unique solution to the differential Lyapunov equation \eqref{eq:lyapunov_bosons} is defined by ~\cite{Behr2019},
\begin{align}\label{eq:solution_lyapunov}
    &\boldsymbol{V}(t) =\, e^{\boldsymbol{A}(t-t_0)}\, \boldsymbol{V}(t_0)\, e^{\boldsymbol{A}^\top (t-t_0)} \nonumber \\
    &\hspace{2.2cm} + \int_{t_0}^t \dd{s} e^{\boldsymbol{A}(t-s)}\, \boldsymbol{D}\, e^{\boldsymbol{A}^\top (t-s)},
\end{align}
where $t_0$ is an initial time for which the covariance matrix is known.
Here, if $\boldsymbol{V}(t_0) \geq 0$ and $\boldsymbol{D} \geq 0$,  then $\boldsymbol{V}(t) \geq 0$ for $t \geq t_0$~\cite{Abou-Kandil2003}; also, since $\boldsymbol{V}(t_0)$ and $\boldsymbol{D} $ are symmetric, the symmetry of $\boldsymbol{V}(t)$ is guaranteed.
Furthermore, if the drift matrix $\boldsymbol{A}$ is Hurwitz (stable), i.e., the real parts of all its eigenvalues are negative, the system will have a unique steady state whose vector of means will be zero ($ \langle \boldsymbol{\x} \rangle_\mathrm{ss} = 0$) and whose covariance matrix $\boldsymbol{V}_\mathrm{ss}$ will be the symmetric positive semidefinite solution to the algebraic Lyapunov equation
\begin{align}\label{eq:algebraic_lyapunov}
   \boldsymbol{A} \boldsymbol{V}_\mathrm{ss} + \boldsymbol{V}_\mathrm{ss} \boldsymbol{A}^\top + \boldsymbol{D} = 0.
\end{align}
The solution to Eq.~\eqref{eq:algebraic_lyapunov} will be given by~\cite{Antsaklis2006},
\begin{align}\label{eq:cm_ss}
    \boldsymbol{V}_\mathrm{ss} = \int_{0}^{\infty} \dd{t} e^{\boldsymbol{A} t}\, \boldsymbol{D}\, e^{\boldsymbol{A}^\top t};
\end{align}
and provided that this unique steady state exists, the solution to Eq.~\eqref{eq:lyapunov_bosons} may be written as
\begin{align}\label{eq:solution_lyapunov_stable}
    \boldsymbol{V}(t) = e^{\boldsymbol{A} t}\left[ \boldsymbol{V}(t_0)-\boldsymbol{V}_\mathrm{ss} \right] e^{\boldsymbol{A}^\top t} + \boldsymbol{V}_\mathrm{ss}.
\end{align}
It is worth noting that for non-Gaussian initial states under linear evolution, Eqs.~\eqref{eq:vector_means_bosons} and \eqref{eq:lyapunov_bosons} still describe the dynamics of their vector of means and covariance matrix.
Further, the non-Gaussianity of the system along with its dynamics can be measured by means of its distinguishability from a reference Gaussian state with the same first and second moments~\cite{Genoni2007,Genoni2008}.
Previously we mentioned that Gaussian states are completely characterized by the vector of means and the covariance matrix.
For bosonic systems, this can be seen in the Gaussian profile of the Wigner function, which will be given in terms of $\langle \boldsymbol{\x} \rangle$ and $\boldsymbol{V}$ by
\begin{align}
    W_\mathrm{G}(\boldsymbol{x}) = \frac{2^N}{\pi^N \sqrt{\mathrm{det}(\boldsymbol{V})}}\,\, e^{-(\boldsymbol{x} - \langle \boldsymbol{\x} \rangle )^\top \boldsymbol{V}^{-1} (\boldsymbol{x} - \langle \boldsymbol{\x} \rangle ) },
\end{align}
where $\boldsymbol{x} =  \left(q_{1}, p_{1}, \dots, q_{N}, p_{N} \right)^\top$~\cite{Olivares2012,Serafini2017}.
On the other hand, it is worth mentioning that using the covariance matrix and the vector of means it is possible to calculate observables of the system such as the zero-delay second-order correlation function $g^{(2)}(0)$~\cite{Olivares2018}.

\subsection{Entanglement of two-mode Gaussian states}
The covariance matrix and the vector of means also allow us to determine entanglement in two-mode CV systems,
where the simplest criterion for entanglement is given by Duan's inequality~\cite{Duan2000}.
Thus, considering the collective quadrature operators
\begin{subequations}\label{eq:duan_0}
\begin{align}
    &\hat{u} = \alpha\, \hat{q}_1 + \beta\, \hat{q}_2, \\
    &\hat{v} = \alpha\, \hat{p}_1 - \beta\, \hat{p}_2,
\end{align}
\end{subequations}
with  $\alpha,\, \beta \in \mathbb{R}$; Duan's criterion establishes that a bosonic state is inseparable if the following inequality is satisfied,
\begin{align}\label{eq:duan}
    \big\langle (\Delta \hat{u})^2 \big\rangle + \big\langle (\Delta \hat{v})^2 \big\rangle < \alpha^2 + \beta^2,
\end{align}
where $\langle (\Delta \O)^2 \rangle = \langle \O^2 \rangle - \langle \O \rangle^2$ is the variance of the Hermitian operator $\O$.
The quantity on the left-hand side of the inequality \eqref{eq:duan}, known as Duan's quantity, can be easily calculated from $\langle \boldsymbol{\x} \rangle$ and $\boldsymbol{V}$.
Duan's criterion is sufficient to determine separability in any bipartite bosonic system; however, for Gaussian CV systems it becomes a necessary and sufficient condition.
Hence, for an entangled Gaussian CV state, it is always possible to find a pair of collective quadrature operators with sufficiently small variance and, therefore, with Einstein-Podolsky-Rosen (EPR) type correlations (see Ref.~\cite{Duan2000} for details).
Furthermore, we can use the covariance matrix and the vector of means to quantify the degree of quantum entanglement of a two-mode Gaussian CV state via the logarithmic negativity \cite{Vidal2002,Audenaert2002,Adesso2004,Plenio2005}.
Thus, if we write the covariance matrix as a block matrix
\begin{align}\label{eq:bosonic_covariance_matrix}
    \renewcommand*{\arraystretch}{1.5}
    \boldsymbol{V}=\left[\begin{array}{c|c}
    \boldsymbol{V}_{1} & \boldsymbol{V}_{1 2} \\
    \hline \boldsymbol{V}_{1 2}^\top & \boldsymbol{V}_{2}
    \end{array}\right],
\end{align}
the logarithmic negativity will be given by
\begin{align}\label{eq:bosonic_logarithmic_negativity}
    \mathcal{E}_{\mathrm{b}} = \max \big\{0,-\ln{\eta} \big\}
\end{align}
where $\quad \eta=2^{-1 / 2} \{ \Sigma( \boldsymbol{V})- [\Sigma(\boldsymbol{V})^{2} - 4\, \mathrm{det} (\boldsymbol{V}) ]^{1 / 2}\}^{1 / 2} \quad$ and
$\Sigma(\boldsymbol{V}) = \mathrm{det} (\boldsymbol{V}_{2}) + \mathrm{det} (\boldsymbol{V}_{1}) - 2\, \mathrm{det} (\boldsymbol{V}_{1 2})$~\cite{Adesso2004,Woolley2014}.
%


\section{Fermionic Gaussian dynamics}
\label{sec:fermionic_linear_dynamics}
Our approach is also valid for fermionic systems, where quantized modes are described by creation and annihilation operators $\c_j^\dagger$ and $\c_j$, satisfying the canonical anti-commutation relation $\{ \c_j, \c_k^\dagger \} = \delta_{j k}$.
These fermionic modes can also be described by Hermitian Majorana operators, given by $\w_{2j-1} = (\c_j^\dagger + \c_j)/\sqrt{2}$, $\w_{2j} = -i (\c_j^\dagger - \c_j)/\sqrt{2}$, which can be collected in an operator-valued vector $\boldsymbol{\w} = (\w_1, \dots, \w_{2N})^\top$, being $N$ the dimension of the Hilbert space.
Majorana operators can be thought of as analogous to position and momentum bosonic operators such that the vector $\boldsymbol{\w}$ defines a phase space for fermionic modes.
In terms of this vector of Majorana operators, the canonical anti-commutation relation takes the form
\begin{align}\label{eq:anticommutation_relation}
    \boldsymbol{\w} \boldsymbol{\w}^\top + \left( \boldsymbol{\w} \boldsymbol{\w}^\top \right)^\top = \mathbb{1},
\end{align}
which in index notation may be written as $\{ \w_j, \w_k \} = \delta_{j k}.$
Similar to the bosonic case, fermionic Gaussian states are fully described by the first and second moments in the Majorana operators according to Wick's theorem~\cite{Wick1950,Bach1994,Botero2004,Corney2006,Surace2022}.
However, for fermionic systems, non-trivial linear combinations of Majorana operators in the Hamiltonian are not physical, so the Liouvillian describing the evolution of the system is always even~($\hat{P} \L \hat{P}^\dagger = \L$, with $\hat{P}$ the parity operator) and, consequently, the density matrix will be even as well ($\hat{P} \rho \hat{P}^\dagger = \rho$). 
Thus, since the vector of Majorana operators is odd ($\hat{P} \boldsymbol{\w} \hat{P}^\dagger = -\boldsymbol{\w}$), the first moments of fermionic states always vanish ($\langle \boldsymbol{\w} \rangle = 0$).
It is clear that this is not the case with bosonic systems, where linear combinations of quadrature operators on the Hamiltonian yield coherent states, which are displaced from the origin in phase space and are not invariant to parity transformations.
Furthermore, the symmetric part of the second moments is fixed by the canonical anti-commutation relation~\eqref{eq:anticommutation_relation} and, therefore, the state of the system will be completely characterized by the antisymmetric part of the second moments in the Majorana operators, which can be collected in the antisymmetric fermionic covariance matrix $\boldsymbol{\sigma} \in\mathbb{R}^{2N \times 2N}$ whose elements are given by $\sigma_{j k} = i \langle [\w_j,\w_k] \rangle$.
A fermionic covariance matrix represents a physical state if and only if it satisfies the positivity condition
\begin{align}\label{eq:positivity_fermions}
    i \boldsymbol{\sigma} \leq \mathbb{1},
\end{align}
which is equivalent to say that the magnitudes of the eigenvalues $\pm i \lambda_j \in i \mathbb{R}$ of $\boldsymbol{\sigma}$ are smaller than or equal to one ($|\lambda_j| \leq 1$)~\cite{Horstmann2013,OnumaKalu2019}.
Moreover, the state is pure if and only if equality is attained ($i \boldsymbol{\sigma} = \mathbb{1}$), which corresponds to $\boldsymbol{\sigma}^2 = -\mathbb{1}$ and $|\lambda_j| = 1$~\cite{Bach1994,Bravyi2005,Surace2022}.
Inequality~\eqref{eq:positivity_fermions} follows from considering the canonical anti-commutation relation~\eqref{eq:anticommutation_relation} along with the positivity of the quantum state $\rho$, and can be interpreted as a consequence of the Pauli exclusion principle~\cite{OnumaKalu2019}.
This condition imposes an additional constraint that is not present in the bosonic case, implying that the space of allowed states is bounded.
In summary, fermionic Gaussian dynamics is significantly more restricted than bosonic Gaussian dynamics.
On the one hand, the vector of means is null and the covariance matrix is antisymmetric, which reduces the degrees of freedom of the system; and on the other hand, the state space will be bounded according to condition~\eqref{eq:positivity_fermions}.
Fermionic Gaussian states play a significant role in quantum information theory~\cite{Bravyi2005,Bravyi2012,deMelo2013}, as well as in the study of out-of-equilibrium quantum many-body systems~\cite{Calabrese2016,Essler2016,Bandyopadhyay2021}.
Moreover, they are remarkably rich in phenomenology, being present for example in models that manifest topological properties such as the Kitaev honeycomb lattice~\cite{Kitaev2006}, and the Su-Schrieffer-Heeger (SSH) model for one-dimensional lattices~\cite{Heeger1988}. 
On the other hand, covariance matrix methods based on the master equation in diagonal Lindblad form have proven useful in the past for the study of non-equilibrium steady states (NESS)~\cite{Prosen2008,Prosen2010},  noise-driven phase transitions in translationally invariant systems \cite{Eisert2010, Horstmann2013}, and decoherence in open Majorana systems \cite{Campbell2015,*Campbell2015b}, just to name a few examples.
For fermionic linear systems, the quadratic Hamiltonian takes the form
\begin{align}\label{eq:hamiltonian_fermions}
    \H = \frac{i}{2} \boldsymbol{\w}^\top \boldsymbol{G}\, \boldsymbol{\w},
\end{align}
which in index notation corresponds to $\H = \frac{i}{2} \sum_{j, k=1}^N G_{j k} \w_j \w_k$, where $\boldsymbol{G}$ is a real antisymmetric matrix.
Further, the decoherence operators will be given by linear combinations of the Majorana operators, 
\begin{align}\label{eq:decoherence_operators_fermions}
    \hat{F}_j = \sum_{k=1}^{2N} M_{j k}\, \w_k, 
\end{align}
where again the $M_{j k}$ are elements of $\boldsymbol{M}\in \mathbb{C}^{M \times 2N}$.
It is worth noting that an $N$-mode fermionic Gaussian state is equivalently defined as a thermal (Gibbs) state, so that its density matrix is an exponential of a quadratic expansion in Majorana operators,
\begin{align}\label{eq:density_matrix_fgs}
    \rho = \frac{1}{Z} e^{-\frac{i}{2} \boldsymbol{\w}^\top \boldsymbol{K}\, \boldsymbol{\w}},
\end{align}
where $\boldsymbol{K}$ is a real antisymmetric matrix and $Z=\Tr[e^{-\frac{i}{2} \boldsymbol{\w}^\top \boldsymbol{K} \boldsymbol{\w}}]$ is a normalization factor. 
From Eq.~\eqref{eq:density_matrix_fgs} it is clear that every fermionic Gaussian state has a normal-mode decomposition given by a quadratic Hamiltonian defined by $\boldsymbol{K}$ [cf. Eq.~\eqref{eq:hamiltonian_fermions}].
Further, the system's covariance matrix will be connected to its density matrix representation via $\boldsymbol{\sigma} = -i \tanh{(i \frac{\boldsymbol{K}}{2})}$~\cite{Kraus2009,Eisler2015,Surace2022}.
Taking into account the anti-commutation relation \eqref{eq:anticommutation_relation} and the antisymmetry of $\boldsymbol{\sigma}$, we can equivalently define the elements of the fermionic covariance matrix as $\sigma_{j k} = 2 i \langle \w_j \w_k \rangle$ for $j \neq k$, and
$\sigma_{j k} = 0$ for $j = k$.
Thus, following a procedure analogous to the one described above for bosonic linear systems, we may write the time-derivative of the elements of $\boldsymbol{\sigma}$ as
\begin{align}\label{eq:derivative_fermionic_covariance}
    \dv{\sigma_{j k}}{t} = 2 i\, \langle \L^\dagger\, \w_j \w_k \rangle, \qquad j\neq k;
\end{align}
with $\L^\dagger$ as defined in Eqs.~(\ref{eq:adjoint_master_equation},\ref{eq:adjoint_master_equation_gl}), being $\H$ and the $\hat{F}_j$ the fermionic versions given in Eqs.~\eqref{eq:hamiltonian_fermions} and \eqref{eq:decoherence_operators_fermions}, respectively.
From Eq.~\eqref{eq:derivative_fermionic_covariance} we find the differential Lyapunov equation for the evolution of the fermionic covariance matrix (see the \hyperref[app:structure_adjoint_equation]{Appendix\ref{app:structure_adjoint_equation}} for more details),
\begin{align}\label{eq:lyapunov_fermions}
    \dv{\boldsymbol{\sigma}}{t} = \boldsymbol{X} \boldsymbol{\sigma} + \boldsymbol{\sigma} \boldsymbol{X}^\top + \boldsymbol{Y},
\end{align}
where the drift matrix $\boldsymbol{X}$ and the diffusion matrix $\boldsymbol{Y}$ are given by
\begin{subequations}
\begin{gather}
    \boldsymbol{X} =  \boldsymbol{G} - \frac{1}{2} \left[  \boldsymbol{M}^\dagger\, \boldsymbol{\Gamma}\, \boldsymbol{M} + (\boldsymbol{M}^\dagger\, \boldsymbol{\Gamma}^\dagger\, \boldsymbol{M})^* \right], \\
    \boldsymbol{Y} = i \left[ \boldsymbol{M}^\dagger\, \boldsymbol{\Gamma}\, \boldsymbol{M} - (\boldsymbol{M}^\dagger\, \boldsymbol{\Gamma}^\dagger\, \boldsymbol{M})^* \right];
\end{gather}
\end{subequations}
with $\boldsymbol{\Gamma}$, $\boldsymbol{G}$, and $\boldsymbol{M}$ \added{as} defined in Eqs.~\eqref{eq:non-diagonal_lindblad_form}, \eqref{eq:hamiltonian_fermions}, and \eqref{eq:decoherence_operators_fermions}, respectively.
Considering the Hermiticity of $\boldsymbol{\Gamma}$, we obtain
\begin{subequations}\label{eq:drift_diffusion_fermions}
\begin{align}
    \boldsymbol{X} &=  \boldsymbol{G} - \Re ( \boldsymbol{M}^\dagger\, \boldsymbol{\Gamma}\, \boldsymbol{M} ), \label{eq:drift_matrix_fermions} \\
    \boldsymbol{Y} &= -2\, \Im ( \boldsymbol{M}^\dagger\, \boldsymbol{\Gamma}\, \boldsymbol{M}). \label{eq:diffusion_matrix_fermions}
\end{align}
\end{subequations}
The dynamical Eq.~\eqref{eq:lyapunov_fermions} together with Eqs.~\eqref{eq:drift_matrix_fermions} and~\eqref{eq:diffusion_matrix_fermions} represent the second part of the main results of this work, which correspond to the equation of motion for the fermionic covariance matrix of a linear system undergoing evolution described by master equation in non-diagonal Lindblad form.
Here, if the master equation describing the system dynamics is in diagonal Lindblad form~\eqref{eq:diagonal_lindblad_form}, we would have as before $\boldsymbol{\Gamma} = \mathrm{diag}( \gamma_1, \dots, \gamma_M) \in \mathbb{R}^{M\times M}$ and $\boldsymbol{C} = \sqrt{\boldsymbol{\Gamma}} \boldsymbol{M}\in \mathbb{C}^{M \times 2N}$.
Therefore, the drift and diffusion matrices in Eqs.~\eqref{eq:drift_diffusion_fermions} would take the known form $\boldsymbol{X} = \boldsymbol{G} - \Re (\boldsymbol{C}^\dagger \boldsymbol{C})$, $\boldsymbol{Y} = - 2\, \Im (\boldsymbol{C}^\dagger \boldsymbol{C})$~\cite{Eisert2010,Bravyi2012}.
The general solution to Eq.~\eqref{eq:lyapunov_fermions} will be of the form \eqref{eq:solution_lyapunov}, where the antisymmetry of the drift matrix $\boldsymbol{Y}$ and the initial condition $\boldsymbol{\sigma}(0)$ ensure that $\boldsymbol{\sigma}(t)$ will be antisymmetric for all $t>0$.
Here, in contrast to the bosonic case, the existence of a stable fixed point of the Liouvillian evolution is guaranteed~\cite{Prosen2008,Prosen2010,Eisert2010} and, therefore, the steady-state covariance matrix $\boldsymbol{\sigma}_\mathrm{ss}$ will be the solution to $\boldsymbol{X} \boldsymbol{\sigma}_\mathrm{ss} + \boldsymbol{\sigma}_\mathrm{ss} \boldsymbol{X}^\top + \boldsymbol{Y} = 0$, which will have the form given in Eq.~\eqref{eq:cm_ss}.
Further, the solution to Eq.~\eqref{eq:lyapunov_fermions} can always be written as shown in Eq.~\eqref{eq:solution_lyapunov}.

As we have seen, the mathematical description of bosonic and fermionic Gaussian states is very similar, however, each one leads to completely different phenomenology.
For a more detailed comparison between bosonic and fermionic Gaussian systems we refer the reader to Refs.~\cite{Eisert2010,Hackl2021,Campbell2015,*Campbell2015b}.
\subsection{Fermionic logarithmic negativity}
For bosonic systems the logarithmic negativity is a versatile and feasible measure of entanglement for Gaussian states~\cite{Audenaert2002,Vidal2002}.
It is defined in terms of the partial transpose of the density matrix, which for such systems is equivalent to a partial time reversal~\cite{Peres1996,Simon2000}.
Unfortunately, for fermionic systems, while partial time reversal corresponds to a Gaussian operator, partial transposition does not~\cite{Eisler2015,Shapourian2017}.
However, it has recently been proven that the definition of fermionic logarithmic negativity based on partial time reversal (fermionic partial transpose) corresponds to a proper measure of entanglement for fermionic systems~\cite{Shapourian2019}.
We take into account here this fermionic logarithmic negativity as it fits very well among the methods based on the covariance matrix.
We shall consider a bipartite Hilbert space $\mathcal{H}= \mathcal{H}_{1} \otimes \mathcal{H}_{2}$, where the fermionic covariance matrix corresponding to the density matrix $\rho$ may be written in block form as
\begin{align}
    \renewcommand*{\arraystretch}{1.5}
    \boldsymbol{\sigma}=\left[\begin{array}{c|c}
    \boldsymbol{\sigma}_{1} &\, \,  \boldsymbol{\sigma}_{1 2} \\
    \hline -\boldsymbol{\sigma}_{1 2}^\top & \boldsymbol{\sigma}_{2}
    \end{array}\right].
\end{align}
The fermionic logarithmic negativity is defined as
\begin{align}
    \mathcal{E}_\mathrm{f} := \ln || \rho^{\R_2} ||_1,
\end{align}
where $|| \O ||_1 = \Tr \sqrt{\O^\dagger \O} $ is the trace norm of the operator $\O$ and $\rho^{\R_2}$ is the partial time-reversal of the density matrix $\rho$ with respect to the second subsystem (see Ref.~\cite{Shapourian2017} for details on the definition of $\rho^{\R_2}$).
For simplicity in the notation, we make $\rho_+ = \rho^{\R_2}$ and $\rho_- = \rho^{\R_2 \dagger}$.
Further, we introduce the auxiliary Gaussian density matrix
\begin{align}
    \rho_\times = \frac{\rho_+ \rho_- }{\Tr( \rho_+ \rho_- )},
\end{align}
whose associated covariance matrix $\boldsymbol{\sigma}_\times$ satisfies
\begin{align}\label{eq:sigmax}
    \boldsymbol{\sigma}_\times \simeq \left( \frac{\mathbb{1} - \boldsymbol{\sigma}^2 }{2} \right)^{-1} \left[\begin{array}{c|c}
     \, \, \boldsymbol{\sigma}_{1}\, \, \,  & \mathbb{0} \\
    \hline \mathbb{0}  & -\boldsymbol{\sigma}_{2}
    \end{array}\right],
\end{align}
where  $\simeq$ represents here the equivalence of the spectra~\cite{Eisert2018,Gruber2020}.
Considering these definitions, the fermionic logarithmic negativity can be written as
\begin{align}\label{eq:fermionic_logarithmic_negativity}
    \mathcal{E}_\mathrm{f} = \frac{1}{2} \left[ S_{1/2}(\rho_\times) - S_2(\rho) \right],
\end{align}
where $S_\alpha(\rho) = \frac{1}{1-\alpha} \ln \Tr(\rho^\alpha) $ is the Rényi entropy of order $\alpha$ ($0<\alpha<\infty$) of the quantum state $\rho$~\cite{Shapourian2017}.
The traces involved in Eq.~\eqref{eq:fermionic_logarithmic_negativity} can be calculated from the spectra $\{\pm i \lambda_j\}$  of $\boldsymbol{\sigma}$ and $\{\pm i \lambda_j^\times\}$ of $\boldsymbol{\sigma}_\times$ according to the following expressions,
\begin{subequations}\label{eq:traces}
\begin{align}
    \Tr(\rho^2) &= \prod_j \left( \frac{1+\lambda_j^2}{2} \right), \\
    \Tr(\rho_\times^{1/2}) &= \prod_j \left[ \left( \frac{1+\lambda_j^\times}{2} \right) + \left( \frac{1-\lambda_j^\times}{2} \right) \right]; 
\end{align}
\end{subequations}
where the double degenerate eigenvalues are to be counted only once~\cite{Eisert2018,Gruber2020}.
Thus, we can calculate the fermionic logarithmic negativity of a bipartite Gaussian system from its covariance matrix using Eq.~\eqref{eq:fermionic_logarithmic_negativity} together with Eqs.~\eqref{eq:sigmax} and~\eqref{eq:traces}.
\subsection{Duan quantity criterion for fermionic systems}
Although Duan's criterion was initially developed as a criterion for entanglement between bosonic systems, it is possible to prove that an analogous version is valid for fermionic systems.
Proceeding from Eq.~(4) in Ref.~\cite{Duan2000}, which is equally valid for fermionic systems if $\hat{u}$ and $\hat{v}$ are defined as collective Majorana operators.
We have then that if $\hat{u}$ and $\hat{v}$ are given by
\begin{subequations}\label{eq:uv_fermions}
\begin{align}
    &\hat{u} = \alpha\, \w_1 + \beta\, \w_3, \\
    &\hat{v} = \alpha\, \w_2 - \beta\, \w_4,
\end{align}
\end{subequations}
with $\alpha,\, \beta \in \mathbb{R}$, and  $\w_{2j-1},\, \w_{2j}$ as defined above; it follows that a fermionic separable bipartite system ($\rho = \sum_i p_i\, \rho_{i_1} \otimes \rho_{i_2}$) will satisfy the inequality
\begin{align}
    &\big\langle (\Delta \hat{u})^2 \big\rangle_\rho + \big\langle (\Delta \hat{v})^2 \big\rangle_\rho \geq \nonumber \\
    &\hspace{1.5cm} \sum_i p_i\, \bigg\{  \alpha^2 \left[ \big\langle (\Delta \w_1)^2 \big\rangle_i + \big\langle (\Delta \w_2)^2 \big\rangle_i \right] \nonumber \\
    &\hspace{2.3cm} + \beta^2 \left[ \big\langle (\Delta \w_3)^2 \big\rangle_i + \big\langle (\Delta \w_4)^2 \big\rangle_i \right] \bigg\},
\end{align}
where $\langle \O \rangle_\rho = \Tr( \rho\, \O) $, $\langle \O \rangle_i = \Tr( \rho_{i_1} \otimes \rho_{i_2}\, \O) $, and $\langle (\Delta \O)^2 \rangle = \langle \O^2 \rangle - \langle \O \rangle^2$ the variance of the Hermitian operator $\O$.
Now, from the Heisenberg-Robertson inequality we have that $\langle (\Delta \hat{A})^2 \rangle + \langle (\Delta \hat{B})^2 \rangle \geq |\langle [\hat{A},\hat{B}] \rangle|$ and, hence, 
\begin{align}\label{eq:fermionic_duan_1}
    &\langle (\Delta \w_{2j-1})^2 \rangle_i + \langle (\Delta \w_{2j})^2 \rangle_i \geq \big|\, 2\, \langle \c_j^\dagger \c_j \rangle_i -1\, \big|,
\end{align}
where $\langle \c_j^\dagger \c_j \rangle_i = \langle 1 | \rho_{i_j} | 1 \rangle$.
It was recently shown by Shapourian and Ryu in Sec. IV A of Ref.~\cite{Shapourian2019} that, from the fermionic logarithmic negativity point of view, a given two-fermion density matrix is separable ($\mathcal{E}_\mathrm{f} = 0 $) if all its off-diagonal elements are zero.
Hence, $\rho$ will be diagonal and, consequently, so will be the $\rho_{i_j}$.
Further, since the latter represent pure states, $\langle \c_j^\dagger \c_j \rangle_i = \langle 1 | \rho_{i_j} | 1 \rangle$ will be either $0$ or $1$ without taking intermediate values, and the right-hand side of inequality \eqref{eq:fermionic_duan_1} will be equal to $1$.
Fermionic separable systems will then satisfy
\begin{align}
    &\langle (\Delta \hat{u})^2 \rangle_\rho + \langle (\Delta \hat{v})^2 \rangle_\rho \geq  \alpha^2 + \beta^2.
\end{align}
It is worth noting that using the uncertainty relation is possible to show that for inseparable states the minimum total variance of the operators $\hat{u}$ and $\hat{v}$ is equal to zero.
The analogue of Duan's criterion for fermionic systems then establishes that a bipartite system is inseparable if the following inequality is satisfied,
\begin{align}\label{eq:duan_fermions}
    \langle (\Delta \hat{u})^2 \rangle + \langle (\Delta \hat{v})^2 \rangle < \alpha^2 + \beta^2,
\end{align}
with $\langle \O \rangle = \langle \O \rangle_\rho $, and $\hat{u}$ and $\hat{v}$ as given in Eqs.~\eqref{eq:uv_fermions}.
Similar to the bosonic case, the quantity on the left-hand side of inequality \eqref{eq:duan_fermions} can be easily calculated from the covariance matrix $\boldsymbol{\sigma}$.
Moreover, this criterion is a sufficient condition to determine the separability of a given bipartite fermionic state.


%
\section{Illustrative example}
\label{sec:ilustrative_examples}
To illustrate the usefulness of the proposed method, we present in this section a physical example in which we study the dissipative entanglement of non-interacting two-mode systems.

\subsection*{Two-mode mechanical system coupled to an engineered high-Q electromagnetic mode}
\begin{figure}[h]
    \centering
    \includegraphics[width=0.7\linewidth]{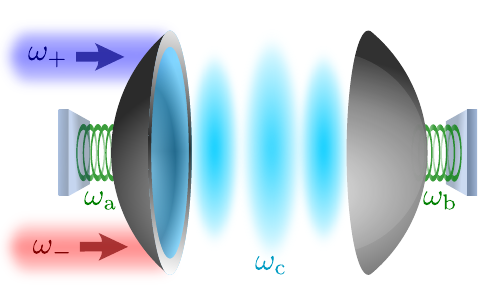}    \caption{Two-mode optomechanical system under two-tone driving. The system consists of two mechanical oscillators, with resonant frequencies $\omega_{\mathrm{a}}$ and $\omega_{\mathrm{b}}$, coupled to a common electromagnetic cavity mode, with resonance frequency $\omega_{\mathrm{c}}$, under two-tone driving. The frequencies of the input tones are $\omega_+$ (blue-shifted) and $\omega_-$ (red-shifted). If the single-photon optomechanical couplings $g_\mathrm{a}$ and $g_\mathrm{b}$ are equal and the input frequencies satisfy $\omega_\pm = \omega_\mathrm{c} \pm  \omega_\mathrm{m}$, with $\omega_\mathrm{m}=(\omega_\mathrm{a} + \omega_\mathrm{b})/2$, one finds the effective Hamiltonian given by Eq.~\eqref{eq:optomechanical_hamiltonian}, where $\omega_1 = -\omega_2 = \Omega = (\omega_\mathrm{a} - \omega_\mathrm{b})/2$ and $G_\pm = (g_\mathrm{a} + g_\mathrm{b}) \bar{c}_\pm/2$, with $\bar{c}_\pm$ the steady-state amplitudes of the input fields. \label{fig:two_mode_squeezing} }
\end{figure}

\begin{figure}[ht!]
    \centering
    \includegraphics[width=0.95\linewidth]{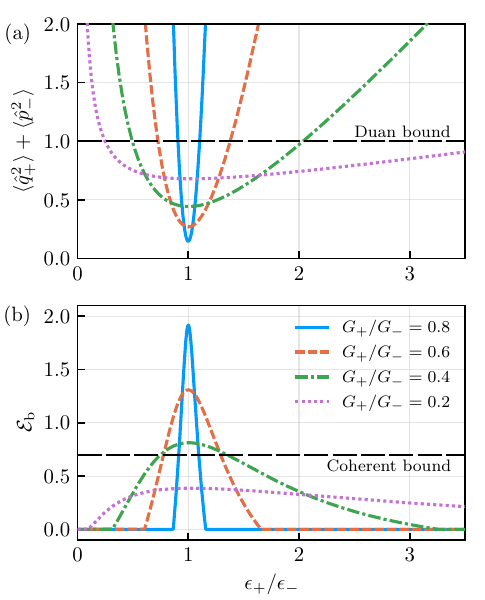}    \caption{Mechanical entanglement, characterized by (a) the Duan criterion [as given by Eq.~\eqref{eq:example_duan}] and (b) the logarithmic negativity $\mathcal{E}_\mathrm{b}$ [calculated from Eq.~\eqref{eq:bosonic_logarithmic_negativity}], as a function of the absorption spectrum asymmetry ($\epsilon_+/\epsilon_-$) for different drive asymmetries ($G_+/G_-$). It is clear that for this example the best scenario takes place when the absorption spectrum is symmetric ($\epsilon_+/\epsilon_-= 1$). Here $\epsilon_+$ was tuned while $\kappa \epsilon_- = 4$,  $G_-/\kappa = 0.2$, $\bar{\gamma}/\kappa = 10^{-4}$, and $\bar{n}=10$. The chosen parameters correspond to a red-tone cooperativity $C_- = 4G_-^2/\bar{\gamma} \kappa = 1600$.  \label{fig:entanglement} }
\end{figure}

%
We shall consider a macroscopic model comprised of a two-mode mechanical system coupled to a high-Q electromagnetic mode under two-tone driving.
The mechanical modes correspond to those of two uncoupled mechanical oscillators interacting simultaneously with a single cavity mode,  which in turn is driven at two sidebands associated with the mechanical motion (see Fig.~\ref{fig:two_mode_squeezing}).
A problem with these characteristics has been studied in the past in the context of two-mode back-action evading (BAE) measurements~\cite{Woolley2013,Ockeloen-Korppi2016,MercierdeLepinay2021} and two-mode squeezing~\cite{Woolley2014,Ockeloen-Korppi2018} in cavity quantum optomechanics.
The effective Hamiltonian for the optomechanical system under consideration reads
\begin{align}\label{eq:optomechanical_hamiltonian}
    &\hat{H} = \sum_{j=1}^2 \left\{ \omega_j\,  \hat{b}^\dagger_j \hat{b}_j +   \left[ ( G_+ \hat{b}_j^\dagger + G_- \hat{b}_j ) \hat{c}^\dagger + \mathrm{H.c.} \right] \right\} + \hat{H}_\mathrm{diss},
\end{align}
where $\hat{b}_j$ is the annihilation operator of the $j$-th mechanical mode, and $\hat{c}$ corresponds to the annihilation operator of the electromagnetic cavity field.
Further, the $G_\pm$ are the many-photon optomechanical couplings, which we will assume real for simplicity; while $\omega_1=-\omega_2=\Omega$, being $\Omega$ an effective oscillator frequency (see the caption of Fig.~\ref{fig:two_mode_squeezing} for the relationship between these quantities and the original system parameters).
The term $\hat{H}_\mathrm{diss}$ accounts for the coupling of the system to Markovian reservoirs, and can be expressed as the sum of a mechanical component $\hat{H}_\mathrm{diss, m}$ and a cavity component $\hat{H}_\mathrm{diss, c}$,
\begin{align}
    \hat{H}_\mathrm{diss} = \hat{H}_\mathrm{diss, m} + \hat{H}_\mathrm{diss, c}.
\end{align}
The Hamiltonian \eqref{eq:optomechanical_hamiltonian} is equivalent to the one given by Eq.~(8) in Ref.~\cite{Woolley2014}, where a detailed derivation can be found in Appendix A of the same reference.
In this example, we adiabatically eliminate the cavity mode, which will leave us with a quantum master equation in non-diagonal Lindblad form.
It is important to note that this adiabatic elimination is valid as long as the relaxation time of the cavity field is much shorter than any time scale associated with the mechanical modes~\cite{Warszawski2000}. 
Being $\kappa$ the cavity bandwidth and taking into account the definitions in the caption of Fig.~\ref{fig:two_mode_squeezing}, this implies that $\kappa > \Omega,\, G_\pm$~\cite{Woolley2014}. 
These conditions are easily satisfied, and the result is a cavity field that adiabatically follows the dynamics of the mechanical modes.
We shall consider the resolved-sideband regime where $\omega_\mathrm{a}, \omega_\mathrm{b} \gg k$ as done in Ref.~\cite{Woolley2014}, however, we will not make a rotating-wave approximation when studying the adiabatic limit. 
Two-mode squeezing generation in the unresolved-sideband regime has recently been studied using an effective master equation in the adiabatic limit in Ref.~\cite{Wang2021}.
We transform now the Hamiltonian~\eqref{eq:optomechanical_hamiltonian} into the interaction picture with respect to $\hat{H}_0 = \sum_{j=1}^2 \omega_j \hat{b}^\dagger_j \hat{b}_j + \hat{H}_\mathrm{diss, c}$, so that the system Hamiltonian may be written as
\begin{align}\label{eq:ex1:hamiltonian}
    \H_{\mathrm{I}}(t) = \c^\dagger(t)\, \hat{f}(t)  + \hat{f}^\dagger(t)\, \c(t),
\end{align}
where $\hat{f}(t)$ is given by
\begin{align}\label{eq:ex1:hamiltonian_1}
    \hat{f}(t) = \sum_{j=1}^2 \Big[ G_+\, e^{i \omega_j t}\, \b_j^\dagger  + G_-\, e^{-i \omega_j t}\, \b_j  \Big];
\end{align}
while $\c(t) = e^{i \hat{H}_\mathrm{diss,c} t}\, \c\, e^{-i \hat{H}_\mathrm{diss,c} t}$, such that the particular structure of $\c(t)$ will depend on the engineering of the cavity field.
By comparing the Hamiltonian described by Eqs.~\eqref{eq:ex1:hamiltonian} and \eqref{eq:ex1:hamiltonian_1} with the general Hamiltonian given by Eqs.~\eqref{eq:interaction_hamiltonian} and \eqref{eq:system_operators}, we can follow the procedure described in Sec.~\ref{sec:reservoir_engineering} to obtain an effective master equation for the mechanical modes alone.
Here, we will not take into account correlations between the mechanical thermal bath and the structured single cavity mode and, accordingly, the effects of $\hat{H}_\mathrm{diss, m}$ on the reduced master equation will be those known from a standard Born-Markov procedure for free damped harmonic oscillators~\cite{Walls2007}.
Thus, under the Born-Markov approximations, the evolution of the reduced density matrix will be given by

\begin{align}\label{eq:ex1:master_equation}
    \dv{\rho(t)}{t} = \left( \mathcal{L}_\mathrm{m} + \mathcal{L}_\mathrm{c}  \right) \rho(t),
\end{align}
where $\mathcal{L}_\mathrm{m}$ accounts for the effects due to $\hat{H}_\mathrm{diss, m}$, while $\mathcal{L}_\mathrm{c}$ describes the interaction with the cavity field.
The Liouvillian superoperator $\mathcal{L}_\mathrm{m}$ takes the form
\begin{align}\label{eq:ex1:mechanical_liouvillian}
    \mathcal{L}_\mathrm{m}\,\rho(t) = \sum_{j=1}^2 \left[ \bar{\gamma}_{j} (\bar{n}_j+1)\, \D[\b_j]\, \rho(t) + \bar{\gamma}_j \bar{n}_j\, \D[\b^\dagger_j]\, \rho(t) \right],
\end{align}
where the $\bar{\gamma}_j$ are the mechanical damping rates, the $\bar{n}_j$ are the Bose-Einstein thermal occupation factors of the respective mechanical reservoirs, and the $\D[\hat{F}_j]$ correspond to the diagonal dissipators defined in Eq.~\eqref{eq:lindblad_dissipator}.
Further, $\mathcal{L}_\mathrm{c}$ can be written as
\begin{align}\label{eq:ex1:master_equation_1}
    \mathcal{L}_\mathrm{c}\, \rho(t) = \sum_{j,k = 1}^{2} & \,\, \left\{ \, \gamma_{j k}^{(1)}\, \left[  \b_j\, \rho(t)\, \b_k^\dagger - \b_k^\dagger \b_j\, \rho(t) \right] \right. \nonumber \\
    & + \gamma_{j k}^{(2)}\, \left[  \b_j\, \rho(t)\, \b_k - \b_k \b_j\, \rho(t) \right] \nonumber \\
    &  + \gamma_{j k}^{(3)}\, \left[  \b_j^\dagger\, \rho(t)\, \b_k^\dagger - \b_k^\dagger \b_j^\dagger\, \rho(t) \right] \nonumber \\
    &  \left. +  \gamma_{j k}^{(4)}\, \left[  \b_j^\dagger\, \rho(t)\, \b_k - \b_k \b_j^\dagger\, \rho(t) \right]\, \right\} + \mathrm{H.c.},
\end{align}
where the decoherence rates are given by
\begin{align}\label{eq:ex1:gammas}
    &\gamma_{j k}^{(m)} = \int_0^\infty \dd{s}\, 
    G_{j k}^{(m)}(s)\, \langle \c(s) \c^\dagger(0) \rangle,
\end{align}
with the $G_{j k}^{(m)}(s)$ defined by 
\begin{subequations}\label{eq:ex1_gs}
\begin{alignat}{2}
    &G^{(1)}_{j k}(s) = G_-^2\, e^{i \omega_j s}, &\qquad& (\omega_j = \omega_k), \\
    &G^{(2)}_{j k}(s) = G_+ G_- \, e^{-i \omega_j s},  &\qquad& (\omega_j = - \omega_k), \\
    &G^{(3)}_{j k}(s) = G_+ G_-\, e^{i \omega_j s}, &\qquad& (\omega_j = - \omega_k), \\
    &G^{(4)}_{j k}(s) = G_+^2\, e^{-i \omega_j s}, &\qquad &  (\omega_j = \omega_k).
\end{alignat}
\end{subequations}
Here we have considered that the correlation function $\langle\c^{\dagger}(s)\c(0)\rangle$ can be neglected. 
This is a fair assumption in the weak-coupling regime, where the backaction of the mechanics on the dynamics of the cavity field is minimal and, consequently, $\langle\c^{\dagger}(s)\c(0)\rangle$ will be directly proportional to the thermal occupancy of the electromagnetic bath.
Furthermore, Eqs.~\eqref{eq:ex1_gs} were obtained following the same logic used to derive Eqs.~\eqref{eq:g_1} and therefore rely on a secular approximation.
The validity of this approximation depends in this case on the fulfillment of the condition $\Omega \gg \bar{\gamma}$, with $\bar{\gamma}=\max(\bar{\gamma}_1, \bar{\gamma}_2)$~\cite{Yamaguchi2017}, which is easily satisfied in optomechanical (electromechanical) systems.
Moreover, the resonance conditions on Eqs.~\eqref{eq:ex1_gs} indicate which decoherence rates will survive, namely $\gamma_{1 1}^{(1)}$, $\gamma_{2 2 }^{(1)}$, $\gamma_{1 2}^{(2)}=\gamma_{2 1}^{(3)}$, $\gamma_{2 1}^{(2)}=\gamma_{1 2}^{(3)}$,  $\gamma_{1 1}^{(4)}$, $\gamma_{2 2}^{(4)}$.
Now, since the $\gamma_{j k}^{(m)}$ are complex numbers, we may express them as $\gamma_{j k}^{(m)} = \Gamma_{j k}^{(m)}/2 + i\Upsilon_{j k}^{(m)}$ and, therefore, Eq.~\eqref{eq:ex1:master_equation_1} takes the form
\begin{align}\label{eq:ex1:master_equation_2}
    \L_{\mathrm{c}}\, \rho(t) =& -i [ \H_\mathrm{LS},\rho(t) ] +  \sum_{j,k=1}^{2} \left\{\, \Gamma_{j k}^{(1)}\, \Ds[\b_j, \b_k]\, \rho(t) \right. \nonumber \\
    &+ \Gamma_{j k}^{(2)}\, \Ds[\b_j, \b_k^\dagger]\, \rho(t) + \Gamma_{j k}^{(3)}\, \Ds[\b_j^\dagger, \b_k]\, \rho(t) \nonumber \\
    & \hspace{2.6cm} \left. + \Gamma_{j k}^{(4)}\, \Ds[\b_j^\dagger, \b_k^\dagger]\, \rho(t)\,  \right\},
\end{align}
where the Lamb shift Hamiltonian is given by
\begin{align}
    &\H_\mathrm{LS} =   \sum_{j,k=1}^{N} \, \left[ \, \Upsilon_{j k}^{(1)} \b_j^\dagger \b_k + \Upsilon_{j k}^{(2)} \b_j^\dagger \b_k^\dagger \right. \nonumber \\ 
    & \left. \hspace{3.4cm} + \Upsilon_{j k}^{(3)} \b_j \b_k + \Upsilon_{j k}^{(4)} \b_j \b_k^\dagger \right],
\end{align}
and the non-diagonal dissipators $\Ds[\hat{F}_j, \hat{F}_k] \rho$ are as defined in Eq.~\eqref{eq:non-diagonal_dissipator}.
Here, we will neglect the Lamb shift Hamiltonian, as the corrections to the energy levels it produces are typically small.
Furthermore, from Eqs.~\eqref{eq:ex1:gammas} and \eqref{eq:ex1_gs} we have that the explicit structure of the non-zero rates $\Gamma_{j k}^{(m)}$ is as follows,
\begin{subequations}\label{eq:ex1_decoherence_rates}
\begin{alignat}{2}
    &\Gamma^{(1)}_{j k} = G_-^2\, S[-\omega_j], &\qquad& (\omega_j = \omega_k), \\
    &\Gamma^{(2)}_{j k} = G_+ G_- \, S[\omega_j],  &\qquad& (\omega_j = - \omega_k), \\
    &\Gamma^{(3)}_{j k} = G_+ G_-\, S[-\omega_j], &\qquad& (\omega_j = - \omega_k), \\
    &\Gamma^{(4)}_{j k} = G_+^2\, S[\omega_j], &\qquad &  (\omega_j = \omega_k);
\end{alignat}
\end{subequations}
where $S[\omega]$ is the cavity absorption spectrum which corresponds to the power spectral density (PSD) of the quantum noise process represented by the operator $\c^\dagger(t)$ (see Appendix B of Ref.~\cite{Bernal-Garcia2020}), and is given by
\begin{align}
    S[\omega] = \int_{-\infty}^{+\infty} \dd{\tau} e^{-i \omega \tau} \langle \c(\tau) \c^\dagger(0) \rangle.
\end{align}
Finally, we define the vector of decoherence operators $\boldsymbol{\hat{F}} = (\b_1, \b_2, \b_1^\dagger, \b_2^\dagger)^\top$ so that using Eqs.~\eqref{eq:ex1:mechanical_liouvillian} and \eqref{eq:ex1:master_equation_2} we may write the master equation \eqref{eq:ex1:master_equation} as 
\begin{align}\label{eq:ex1:master_equation_3}
    \dv{\rho(t)}{t} = \sum\limits_{j, k=1}^{4} \Gamma_{j k}\, \Ds[ \hat{F}_j, \hat{F}_k ]\, \rho(t),
\end{align}
where the $\Gamma_{j k}$ are elements of the decoherence matrix $\boldsymbol{\Gamma}$, which takes the form

\begin{align}\label{eq:example_decoherence_matrix}
\boldsymbol{\Gamma} = \left(\begin{array}{cccc}
    \Gamma^{(1)}_{11} + \bar{\Gamma}^{(1)}_1 & 0 & 0 & \Gamma^{(2)}_{12} \\
    0 & \Gamma^{(1)}_{22}  + \bar{\Gamma}^{(1)}_2 & \Gamma^{(2)}_{21} & 0 \\
    0 & \Gamma^{(2)}_{21} & \Gamma^{(4)}_{11} + \bar{\Gamma}^{(4)}_1 & 0 \\
    \Gamma^{(2)}_{12} & 0 & 0 & \Gamma^{(4)}_{22}  + \bar{\Gamma}^{(4)}_2
    \end{array}\right)
\end{align}
with $\bar{\Gamma}^{(1)}_{j} = \bar{\gamma}_{j} (\bar{n}_j+1)$, $\bar{\Gamma}^{(4)}_{j} = \bar{\gamma}_j \bar{n}_j$.
The non-zero elements of $\boldsymbol{\Gamma}$ are then given by
\begin{subequations}
\begin{align}
    \Gamma^{(1)}_{11} + \bar{\Gamma}^{(1)}_1 &= G_-^2 S[-\Omega] + \bar{\gamma}_{1} ( \bar{n}_1+1), \\
    \Gamma^{(1)}_{22} + \bar{\Gamma}^{(1)}_2 &= G_-^2 S[\Omega] + \bar{\gamma}_{2} (\bar{n}_2+1), \\
    \Gamma^{(4)}_{11} + \bar{\Gamma}^{(4)}_1 &= G_+^2 S[\Omega] + \bar{\gamma}_{1} \bar{n}_1,\\
    \Gamma^{(4)}_{22} + \bar{\Gamma}^{(4)}_2 &= G_+^2 S[-\Omega] + \bar{\gamma}_{2} \bar{n}_2,\\
    \Gamma^{(2)}_{12}  &= G_+ G_- S[\Omega],\\
    \Gamma^{(2)}_{21}  &= G_+ G_- S[-\Omega].
\end{align}
\end{subequations}
Next, we are interested in the steady-state solution of the covariance matrix $\boldsymbol{V}$, in order to study the degree of entanglement between mechanical modes.
To do so, we shall solve the algebraic Lyapunov equation \eqref{eq:algebraic_lyapunov} with the drift and diffusion matrices as defined in Eqs.~\eqref{eq:drift_diffusion_bosons}.
It is important to note that using these equations requires writing the decoherence matrix in the basis of quadrature operators $\boldsymbol{\x} = \left(\hat{q}_{1}, \hat{p}_{1}, \hat{q}_{2}, \hat{p}_{2} \right)^\top$, which can be done through the transformation matrix $\boldsymbol{M}$ defined in Eq.~\eqref{eq:decoherence_operators} and here given by
\begin{align}\label{eq:example_transformation_matrix}
\boldsymbol{M} = \frac{1}{\sqrt{2}} \left(\begin{array}{cccc}
    1 & i & 0 & 0 \\
    0 & 0 & 1 & i \\
    1 & -i & 0 & 0 \\
    0 & 0 & 1 & -i
    \end{array}\right).
\end{align}
Therefore, with the matrices $\boldsymbol{\Omega}$, $\boldsymbol{M}$ and $\boldsymbol{\Gamma}$ as defined in Eqs.~\eqref{eq:symplectic_form}, \eqref{eq:example_transformation_matrix}, \eqref{eq:example_decoherence_matrix}, respectively, we can use Eq.~\eqref{eq:drift_matrix_bosons} to determine the drift matrix $\boldsymbol{A}$ and Eq.~\eqref{eq:diffusion_matrix_bosons} for the diffusion matrix $\boldsymbol{A}$.
Given that we are working in the interaction picture and have neglected the Lamb shift Hamiltonian, there is no Hamiltonian matrix $\boldsymbol{H}$ in this specific example.
While we will not present the explicit form of $\boldsymbol{A}$ and $\boldsymbol{D}$ here, it is important to note that they share the same non-zero structure as the $\boldsymbol{\Gamma}$ matrix in Eq.~\eqref{eq:example_decoherence_matrix}, and can be brought into a block diagonal form with the help of the unitary transformation $\boldsymbol{M}$.
Operating with these matrices in block diagonal form greatly simplifies the solution of the algebraic Lyapunov equation \eqref{eq:algebraic_lyapunov}, since the problem is reduced to the solution of two (complex) Lyapunov equations in $\mathbb{C}^{2\times 2}$.
Furthermore, before proceeding with the characterization of entanglement in this mechanical bipartite system, it is pertinent to recognize that due to the non-singularity of the obtained matrix $\boldsymbol{A}$, the vector of means will be zero in the steady state [cf.~Eq.~\eqref{eq:vector_means_bosons}].
We may express the Duan criterion \eqref{eq:duan} in terms of the collective mechanical quadratures $\hat{q}_\pm$ and $\hat{p}_\pm$ defined as 
\begin{subequations}
\begin{align}
    &\hat{q}_\pm =  \frac{1}{\sqrt{2}} \left(\hat{q}_1 \pm \hat{q}_2 \right), \\
    &\hat{p}_\pm =  \frac{1}{\sqrt{2}} \left(\hat{p}_1 \pm \hat{p}_2 \right);
\end{align}
\end{subequations}
so that we may choose $\alpha = \beta = 1/\sqrt{2}$ in Eqs.~\eqref{eq:duan_0} and \eqref{eq:duan}, and the mechanical state will be inseparable if the following inequality is satisfied,
\begin{align}\label{eq:example_duan}
    \big\langle (\Delta \hat{q}_+)^2 \big\rangle + \big\langle (\Delta \hat{p}_- )^2 \big\rangle < 1,
\end{align}
Thus, the Duan quantity will be given by [with the covariance matrix arranged as shown in Eq.~\eqref{eq:bosonic_covariance_matrix}]
\begin{align}
    &\big\langle (\Delta \hat{q}_+)^2 \big\rangle + \big\langle (\Delta \hat{p}_- )^2 \big\rangle = \big\langle \hat{q}_+^2 \big\rangle + \big\langle \hat{p}_-^2 \big\rangle \nonumber \\
    & \hspace{1cm} = \frac{1}{4} \left[ V_{11} + V_{22} + V_{33} + V_{44} + 2 V_{13} - 2 V_{24} \right],
\end{align}
where the steady state solution of the covariance matrix is obtained as described above.
The Duan quantity as a function of the system parameters is plotted in Fig~\ref{fig:entanglement}.
It is important to emphasize that using the described method it is possible to obtain an analytical solution to the covariance matrix in the steady state, from which we obtain the following expression for the variances of the collective operators,
\begin{widetext}
\begin{align}\label{eq:ex:expectation_values}
    \langle \hat{q}_{\pm}^2 \rangle = \langle \hat{p}_{\mp}^2 \rangle = \frac{ \left( \splitfrac{  (\C_- - \C_+)^2\,  e^{\pm 2 r}\, \left[\,  (\C_-+\C_+) \left( e^{\pm 4 r} - 2 \right)   + 4\, ( \bar{n}+1/2)\, e^{\pm 2 r} - 4 \, \right]  } { + \left[\,  (\C_-+\C_+)\, e^{\mp 2 r} + 4\, (\bar{n} +1/2) \, \right]
   \left[\, \left(\C_-^2 + 6\, \C_- \C_+ + \C_+^2 \right)  + 8 \, (\C_-+\C_+) + 8 \, \right] } \right) }{16 \left[\, \C_-+\C_+ +2 \, \right] \left( \C_- + 1 \right) \left( \C_+ + 1\right) },
\end{align}
\end{widetext}
where we have reparametrized the problem in terms of the effective optomechanical coupling $\G$ and the two-mode squeezing parameter $r$, defined as
\begin{subequations}
    \begin{align}
    &\G^2 = G_-^2 - G_+^2,\\
    &\tanh{r} = \frac{G_+}{G_-},
\end{align}
\end{subequations}
such that $0<G_+/G_+<1$.
Further, the cooperativities $\C_\pm$ are defined by
\begin{align}
    \C_\pm=\frac{\G^2 \epsilon_\pm }{\bar{\gamma}};
\end{align}
where for simplicity we have made
\begin{align}
    \epsilon_\pm = S[\pm\Omega],
\end{align}
and have assumed that the mechanical damping rates ($\bar{\gamma}_j$) and the mean number of thermal phonons in the corresponding mechanical baths ($\bar{n}_j$) are equal for both oscillators,
\begin{subequations}
\begin{align}
    &\bar{\gamma} = \bar{\gamma}_1 = \bar{\gamma}_2, \\
    &\bar{n} = \bar{n}_1 = \bar{n}_2.
\end{align}
\end{subequations}
Equation~\eqref{eq:ex:expectation_values} reduces to Eq.~(29a) of Ref.~\cite{Woolley2014} in the limit of a symmetric absorption spectrum ($\epsilon_- = \epsilon_+$).
In summary, engineering the dissipative dynamics of the system can be done through modification of the structure of the cavity absorption spectrum $S[\omega]$, where asymmetry can be achieved using auxiliary electromagnetic modes, for example, as proposed in Refs.~\cite{Zhang2019,Wang2021}.
Furthermore, using Eq.~\eqref{eq:bosonic_logarithmic_negativity} we can calculate the corresponding logarithmic negativity of the mechanical collective state $\mathcal{E}_\mathrm{b}$, which is shown in Fig.~\ref{fig:entanglement}.
It should be noted that, even though the effect of asymmetry in the cavity absorption spectrum has not been considered before, it is clear that the achieved mechanical entanglement is optimized when the absorption spectrum is symmetric ($\epsilon_+ = \epsilon_-$).

\section{Conclusions}
\label{sec:conclusions}
We have considered here systems described by master equations in a non-diagonal Lindblad form.
This type of master equations arises naturally in quantum reservoir engineering, and although it is in general possible to cast them into a diagonal form by means of a diagonalization procedure, this is often an inefficient way of solving the system dynamics, and it is particularly detrimental when seeking analytical solutions.
For this reason, we have considered here a method to describe the dynamics of engineered quantum systems directly from the master equation in non-diagonal form.
Thus, we have presented a set of evolution equations for the first and second moments of the canonical variables, quadrature operators for bosonic systems and Majorana operators for fermionic systems that allow one to fully characterize the dynamics of linear quantum systems described by master equations in non-diagonal Lindblad form.
Furthermore, using these equations it is possible to obtain exact analytical solutions for the steady state, as is shown in a practical example regarding two-mode mechanical entanglement in an optomechanical system.
This example not only demonstrates the utility of our methods but also indicates their potential for unveiling new phenomena in unexplored physical regimes.
As a review, we have included some covariance matrix methods pertinent to the study of linear open quantum systems.
In particular, we focused on entanglement measures, such as the Duan criterion and logarithmic negativity, and their application to both bosonic and fermionic systems.
For bosonic systems, we revisited established formulas for computing these entanglement measures. For fermionic systems, we highlighted recent advances in calculating the fermionic logarithmic negativity.
A significant outcome of our study is the confirmation that the Duan criterion, traditionally applied to bosonic systems, holds true for fermionic systems as well, as demonstrated by the fermionic logarithmic negativity.
%

\section*{Acknowledgments}
D.N.B.-G. and M.J.W. acknowledge support from ARC Centre for Engineered Quantum Systems (CE170100009) and AFOSR FA 2386-18-1-4026 through the Asian Office of Aerospace Research and Development (AOARD).
D.N.B.-G. acknowledges the support of Colciencias (now Minciencias) through “Convocatoria No. 727 de 2015 – Doctorados Nacionales 2015”. 
D.N.B.-G. gratefully acknowledges financial support from the
Australian Government via AUSMURI Grant No. AUSMURI000002.
L.H. and A.E.M. sincerely thank the support from the Australian Research Council Discovery Project (DP200101353).
A.E.M. was supported by the UNSW Scientia Fellowship program.
%

%
\bigskip
\appendix*
\section{Structure of the adjoint master equation for linear quantum systems}
\label{app:structure_adjoint_equation}
%
%

In this appendix, we discuss the structure of the adjoint master equation in non-diagonal Lindblad form for linear bosonic and fermionic systems.
Knowing the form of the corresponding adjoint superoperators considerably facilitates the derivation of the moment evolution equations described in the main text.

\subsection{Bosons}

From the definition of the adjoint Liouvillian in Eqs.~\eqref{eq:adjoint_master_equation} and \eqref{eq:adjoint_master_equation_gl}, we have that an arbitrary system operator $\O$ satisfies,
\begin{align}\label{eq:app:adjoint_equation}
    &\dv{ \O }{t} = \mathcal{L}^\dagger \O  =\, i [\H, \O] \nonumber \\
    & \hspace{0.3cm} + \sum\limits_{j,k=1}^M \Gamma_{j k} \left[ \hat{F}_k^\dagger \O \hat{F}_j - \frac{1}{2} \left( \O \hat{F}_k^\dagger \hat{F}_j + \hat{F}_k^\dagger \hat{F}_j \O \right) \right], 
\end{align}
where $\boldsymbol{\Gamma}$ is the decoherence matrix and the $\hat{F}_j$ are the decoherence operators.
Further, if we are considering a bosonic linear system, from Eqs.~\eqref{eq:hamiltonian_bosons} and \eqref{eq:decoherence_operators} we know that the Hamiltonian and the decoherence operators may be written in terms of quadrature operators as $\H = \frac{1}{2} \sum_{j,k = 1}^N H_{j k}\, \x_j \x_k$, $\hat{F}_j = \sum_{k=1}^{2N} M_{j k}\, \x_k$.
Therefore, Eq.~\eqref{eq:app:adjoint_equation} takes the form
\begin{align}\label{eq:adjoint_bosons}
    &\dv{ \O }{t} = \mathcal{L}^\dagger \O = \frac{i}{2} \sum_{j,k=1}^N H_{jk}[\x_j \x_k, \O] \nonumber \\
    & + \sum\limits_{j,k=1}^M \sum_{m,n=1}^{2 N} \Gamma_{j k} M_{j n} M_{k m}^* \left[ \x_m\, \O\, \x_n - \frac{1}{2} \left\{ \x_m \x_n , \O \right\}  \right],
\end{align}
where we used $\hat{F}_j^\dagger = \sum_{k=1}^{2N} M_{j k}^*\, \x_k$.
Equation~\eqref{eq:adjoint_bosons} gives the structure of the adjoint Liouvillian for a bosonic linear system in terms of the quadrature operators and the matrices $\boldsymbol{H}$, $\boldsymbol{\Gamma}$, and $\boldsymbol{M}$ that describe the system.
Using this result we can now calculate the moment evolution equations \eqref{eq:vector_means_bosons} and \eqref{eq:lyapunov_bosons}, with the definitions for the drift and diffusion matrices as given in Eqs.~\eqref{eq:drift_diffusion_bosons}, by taking into account that
\begin{subequations}
\begin{gather}
    \dv{\langle \hat{x}_j \rangle}{t} =  \Tr \left[ 
    \left( \mathcal{L}^\dagger \hat{x}_j  \right) \rho \right] = \langle \mathcal{L}^\dagger \hat{x}_j \rangle, \\  
    \dv{ V_{j k} }{t} =  \langle \L^\dagger (\x_j \x_k + \x_k \x_j) \rangle - 2 \dv{}{t} \left( \langle \x_j \rangle \langle \x_k \rangle \right);
\end{gather}
\end{subequations}
being $V_{j k}$ the elements of the covariance matrix $\boldsymbol{V}$.
In the derivations, it will be important to consider the canonical commutation relation $\left[\x_{j}, \x_{k}\right]= i \Omega_{j k}$, with $\boldsymbol{\Omega}$ as given in Eq.~\eqref{eq:symplectic_form}.
\subsection{Fermions}
Similarly to the above, from Eqs.~\eqref{eq:hamiltonian_fermions} and \eqref{eq:decoherence_operators_fermions} we have that for linear fermionic systems the Hamiltonian and decoherence operators may be expressed in terms of Majorana operators as $\H = \frac{i}{2} \sum_{j, k=1}^N G_{j k} \w_j \w_k$, $\hat{F}_j = \sum_{k=1}^{2N} M_{j k}\, \w_k$.
This yields that the adjoint master equation will take the form
\begin{align}\label{eq:adjoint_fermions}
    &\dv{ \O }{t} = \mathcal{L}^\dagger \O = -\frac{1}{2} \sum_{j,k=1}^N G_{jk}[\w_j \w_k, \O] \nonumber \\
    & + \sum\limits_{j,k=1}^M \sum_{m,n=1}^{2 N}  \Gamma_{j k}  M_{j n} M_{k m}^* \left[ \w_m\, \O\, \w_n - \frac{1}{2} \left\{ \w_m \w_n , \O \right\}  \right]. 
\end{align}
Using this expression is not difficult to find the evolution equation for the fermionic covariance matrix $\boldsymbol{\sigma}$ from Eq.~\eqref{eq:derivative_fermionic_covariance},
\begin{align}
    \dv{\sigma_{j k}}{t} = 2 i\, \langle \L^\dagger\, \w_j \w_k \rangle, \qquad j\neq k.
\end{align}
Here, in order to complete the derivation of Eq.~\eqref{eq:lyapunov_fermions}, together with the definitions in Eqs.~\eqref{eq:drift_diffusion_fermions}, it will be necessary to take into account the canonical anti-commutation relation $\{ \w_j, \w_k \} = \delta_{j k}$.
%

%
\bibliography{ms.bib}
%
%

\end{document}